\documentclass[reprint,aps,pra,notitlepage,superscriptaddress]{revtex4-1}

\usepackage{ifthen}
\newboolean{FinalVersion}
\setboolean{FinalVersion}{true}
\newboolean{JournalVersion}
\setboolean{JournalVersion}{true}
\newboolean{IncludeSI}
\setboolean{IncludeSI}{true}

\ifthenelse{\boolean{JournalVersion}}{
  
}{
    \usepackage{mathpazo}
}

\ifthenelse{\boolean{FinalVersion}}{

}{

}

\usepackage{booktabs} 
\usepackage{dcolumn}
\usepackage{bm}

\usepackage[normalem]{ulem}

\usepackage[english]{babel}
\usepackage{colonequals}
\usepackage[colorlinks]{hyperref}
\usepackage{hypcap}
\usepackage{enumitem}

\usepackage{amsmath}
\usepackage{amsfonts,amssymb,amsthm, bbm,braket}
\usepackage{graphicx}   
\usepackage{amsbsy} 
\usepackage{accents}
\usepackage[bold]{hhtensor}

\usepackage{algpseudocode}

\usepackage{multirow}
\usepackage{subcaption}
\DeclareCaptionType{algorithm}
\DeclareCaptionSubType{algorithm}

\setlist[enumerate]{itemsep=0mm}

\newcommand{\nn}{\nonumber\\}

\newcommand{\tA}[1]{\tilde{A}^{(#1)}}
\newcommand{\ketbra}[2]{\left|#1\right\rangle\!\left\langle #2 \right|}

\newcommand{\defeq}{\colonequals}
\newcommand{\suppress}[1]{}

\def\squareforqed{\hbox{\rlap{$\sqcap$}$\sqcup$}}
\def\qed{\ifmmode\squareforqed\else{\unskip\nobreak\hfil
\penalty50\hskip1em\null\nobreak\hfil\squareforqed
\parfillskip=0pt\finalhyphendemerits=0\endgraf}\fi}

\newcommand{\eq}[1]{\hyperref[eq:#1]{(\ref*{eq:#1})}}
\renewcommand{\sec}[1]{\hyperref[sec:#1]{Section~\ref*{sec:#1}}}
\newcommand{\app}[1]{\hyperref[app:#1]{Appendix~\ref*{app:#1}}}
\newcommand{\tab}[1]{\hyperref[tab:#1]{Table~\ref*{tab:#1}}}
\newcommand{\fig}[1]{\hyperref[fig:#1]{Figure~\ref*{fig:#1}}}
\newcommand{\figx}[2]{\hyperref[fig:#1]{Figure~\ref*{fig:#1}(#2)}}
\newcommand{\thm}[1]{\hyperref[thm:#1]{Theorem~\ref*{thm:#1}}}
\newcommand{\lem}[1]{\hyperref[lem:#1]{Lemma~\ref*{lem:#1}}}
\newcommand{\cor}[1]{\hyperref[cor:#1]{Corollary~\ref*{cor:#1}}}
\newcommand{\defn}[1]{\hyperref[def:#1]{Definition~\ref*{def:#1}}}
\newcommand{\alg}[1]{\hyperref[alg:#1]{Algorithm~\ref*{alg:#1}}}

\DeclareMathOperator{\Cov}{Cov}

\newcommand{\T}{\mathrm{T}}
\newcommand{\id}{\mathbbm{1}}
\renewcommand{\openone}{\id}
\newcommand{\expect}{\mathbb{E}}

\newcommand{\Hinv}{H_{-}}

\renewcommand{\H}{\mathcal{H}}

\newcommand{\Tr}{\operatorname{Tr}}

\newcommand{\swapgt}{\textsc{swap}}

\newcommand{\dd}{\mathrm{d}}

\newcommand{\Hint}{\ensuremath{H_{\mathrm{int}}}}
\newcommand{\Hintcap}{\ensuremath{H_{\mathrm{int}\bigcap A}}}
\newcommand{\Hintminus}{\ensuremath{H_{\mathrm{int}\setminus A}}}
\newcommand{\Hout}{\ensuremath{H_{\mathrm{out}}}}
\newcommand{\Hin}{\ensuremath{H_{\mathrm{in}}}}

\makeatletter
\def\Decl@Mn@Delim#1#2#3#4{%
  \if\relax\noexpand#1%
    \let#1\undefined
  \fi
  \DeclareMathDelimiter{#1}{#2}{#3}{#4}{#3}{#4}}
\def\Decl@Mn@Open#1#2#3{\Decl@Mn@Delim{#1}{\mathopen}{#2}{#3}}
\def\Decl@Mn@Close#1#2#3{\Decl@Mn@Delim{#1}{\mathclose}{#2}{#3}}
\DeclareFontFamily{OMX}{MnSymbolE}{}
\DeclareFontShape{OMX}{MnSymbolE}{m}{n}{
    <-6>  MnSymbolE5
   <6-7>  MnSymbolE6
   <7-8>  MnSymbolE7
   <8-9>  MnSymbolE8
   <9-10> MnSymbolE9
  <10-12> MnSymbolE10
  <12->   MnSymbolE12}{}
\DeclareFontShape{OMX}{MnSymbolE}{b}{n}{
    <-6>  MnSymbolE-Bold5
   <6-7>  MnSymbolE-Bold6
   <7-8>  MnSymbolE-Bold7
   <8-9>  MnSymbolE-Bold8
   <9-10> MnSymbolE-Bold9
  <10-12> MnSymbolE-Bold10
  <12->   MnSymbolE-Bold12}{}
\DeclareSymbolFont{mnsymbols}  {OMX}{MnSymbolE}{m}{n}
\Decl@Mn@Open {\lsem}               {mnsymbols}{'102}
\Decl@Mn@Close{\rsem}               {mnsymbols}{'107}
\Decl@Mn@Open {\llangle}            {mnsymbols}{'164}
\Decl@Mn@Close{\rrangle}            {mnsymbols}{'171}
\makeatother

\newlength{\dhatheight}

\begin{document}
\title{Quantum Bootstrapping via Compressed Quantum Hamiltonian Learning}
\author{Nathan Wiebe}
\email{nawiebe@microsoft.com}
\affiliation{Quantum Architectures and Computation Group, Microsoft Research, Redmond, WA 98052, USA}

\author{Christopher Granade}
\email{cgranade@cgranade.com}
\affiliation{Department of Physics, University of Waterloo, Ontario N2L 3G1, Canada}
\affiliation{Institute for Quantum Computing, University of Waterloo, Ontario N2L 3G1, Canada}

\author{D. G. Cory}
\affiliation{Department of Chemistry, University of Waterloo, Ontario N2L 3G1, Canada}
\affiliation{Institute for Quantum Computing, University of Waterloo, Ontario N2L 3G1, Canada}
\affiliation{Perimeter Institute, University of Waterloo, Ontario N2L 2Y5, Canada}
\affiliation{Canadian Institute for Advanced Research, Toronto, Ontario M5G 1Z8}

\begin{abstract}
A major problem facing the development of quantum computers or large scale quantum simulators is that general methods for characterizing and controlling are intractable.   We
provide a new approach to this problem that uses small quantum simulators to efficiently characterize and learn
control models for larger devices.
Our protocol achieves this by using Bayesian inference in concert with Lieb-Robinson bounds and
interactive quantum learning methods to achieve compressed simulations for characterization. 
We also show that the Lieb--Robinson velocity is epistemic for our protocol, meaning that
information propagates at a rate that depends on the uncertainty in the system Hamiltonian.
We
illustrate the efficiency of our bootstrapping protocol by showing numerically that an 8-qubit Ising
model simulator can be used to calibrate and control a 50 qubit Ising simulator while using only about
750 kilobits of experimental data.  Finally, we provide upper bounds for the Fisher information that show that the number of experiments needed to characterize a system rapidly diverges as the duration of the experiments used in the characterization shrinks, which motivates the use of methods such as ours that do not require short evolution times.
\end{abstract}

\maketitle

%





Rapid progress has been made within the last few years towards building computationally useful quantum simulators or computers, which
promise to revolutionize the ways in which we solve problems in chemistry and
material science, data analysis and
cryptography \cite{wiebe_quantum_2012,kassal_simulating_2011,hastings_improving_2014,shor_polynomial-time_1995,amento_efficient_2013}.
Despite this, looming challenges involving
calibrating and debugging quantum devices suggests another possible application for
a small scale quantum computer: designing a larger quantum computer.
This application is increasingly relevant as experiments push towards
building  fault-tolerant devices \cite{chow_implementing_2014}
and demonstrating large scale verifiable quantum computing protocols \cite{barz_demonstration_2012}.

This task can be quite challenging classically.  Simply characterizing the Hamiltonian dynamics of the system via tomography is inefficient~\cite{gross2010quantum} and existing efficient methods such as~\cite{da_silva_practical_2011} require an amount of data that
scales polynomially with the error tolerance, are not known to be error robust and are only efficient for specific classes of Hamiltonians and measurements.  This can render them impractical for problems like designing controls for quantum systems where exacting error tolerance and low fault sensitivity is required.  Other methods are error robust and use a logarithmic amount of data, but also require performing quantum simulations that are intractable classically~\cite{granade_robust_2012,stenberg_efficient_2014}.  The use of quantum simulators has been proposed
as a solution to this problem~\cite{wiebe_hamiltonian_2014,wiebe_quantum_2014-1} but such schemes do not provide a means for characterizing and controlling a large quantum system because they require a 
simulator that is~\emph{at least as large} as the system being characterized.  
{Other schemes have been introduced that allow a small quantum system to efficiently \emph{certify} that a large quantum system implements a set of quantum gates or prepares a given state~\cite{barz_demonstration_2012,aharonov2008interactive,kapourniotis2014verified,aolita2014reliable}.  These schemes are inspired by multi--prover systems and output a certificate that states whether the errors in the gates are above or below a threshold.  Such protocols are rigorous, require very weak assumptions about the errors in the larger system and allow the verifier to use a rudimentary quantum device. Unfortunately, they can also be computationally expensive and are difficult to apply to the case of certifying analog simulation. In particular, existing certification schemes that use small quantum verifiers do not \emph{characterize} the larger system's Hamiltonian dynamics.}


We provide a framework in this paper for overcoming the aforementioned obstacles {to quantum device charactization}.  The idea behind our approach is to use a small simulator as a point of reference
to compare the dynamics of the larger system against.  We achieve this by using an interactive protocol wherein the dynamics of subsystems of the larger device
are measured against the dynamics of the smaller system; thereby allowing us to infer a model for the larger system using the data collected about the
relative dynamics of the two models on the individual subsystems.  We call this process compressed quantum Hamiltonian learning.

It is often insufficient to give a model of a system relative to an
experimental device for which no mathematical model is known. To this end, we
consider starting with a small quantum simulator for which a
firm mathematical model is known.  This system, which we call the \emph{trusted
simulator}~\cite{wiebe_hamiltonian_2014,wiebe_quantum_2014-1} is the key to
allowing our method to provide an absolute, rather than a relative, model for
the larger system's quantum dynamics.  

Compressed Hamiltonian learning in turn leads to the ability to learn models
of \emph{control} dynamics, rather than simply the internal dynamics intrinsic
to the untrusted device under study. This inferred model for control dynamics
then allows compressed quantum Hamiltonian learning to be performed again, with the
previously-untrusted system now being used as the trusted simulator. That is,
compressed quantum Hamiltonian learning directly enables \emph{quantum bootstrapping}, the process of
iteratively building larger trusted simulators out of smaller trusted simulators.

To summarize, we provide two distinct applications:
\setlist[description]{font=\normalfont\itshape\space}
\begin{description}
\item[Compressed QHL (cQHL)]  Learning a Hamiltonian model for a large quantum system using a small quantum simulator.
\item[Quantum bootstrapping] Designing and calibrating controls for a quantum system with rapidly decaying interactions using a smaller quantum simulator.
\end{description}
{These applications are efficient if the unknown Hamiltonian belongs to an efficiently parameterized class of model Hamiltonians for which the interaction strengths between subsystems
decay rapidly with distance, experiments are chosen such that the resultant distribution of likelihoods is far from uniform and that the control maps in the bootstrapping case are well conditioned.  Efficiency also requires that the resampling algorithms used in the methods 
do not introduce substantial error in the system.  This assumption empirically holds for non--degenerate learning problems, such as those we consider here and in prior work~\cite{wiebe_hamiltonian_2014,wiebe_quantum_2014-1}.}

The remainder of the paper is laid out as follows.  We first provide a review
of Bayesian methods for Hamiltonian characterization and Quantum Hamiltonian
Learning in~\sec{review}.  We then discuss compressed quantum Hamiltonian
learning in~\sec{QHL} and quantum bootstrapping in~\sec{bootstrapping}.  We then present
in~\sec{numerical} numerical results for bootstrapping and compressed quantum
Hamiltonian in an important special case, characterizing and controlling $50$
qubit Ising model simulators  before concluding.

\section{Review of Bayesian characterization and quantum Hamiltonian learning}
\label{sec:review}

In developing compressed quantum Hamiltonian learning, and hence quantum
bootstrapping, we will use Bayesian particle filters as
a subroutine. Here, we briefly review these methods in a classical context,
as well as with the inclusion of quantum resources.

\subsection{Bayesian Characterization and Sequential Monte Carlo}

Bayesian methods have been used in a wide range of quantum information applications and experiments;
for instance, to discriminate \cite{becerra_experimental_2013} or estimate states
\cite{blume-kohout_optimal_2010,huszar_adaptive_2012},
to incorporate models of noisy measurements \cite{ferrie_estimating_2012},
to characterize drifting frequencies \cite{shulman_suppressing_2014}, and to estimate
Hamiltonians \cite{sergeevich_characterization_2011,schirmer_quantum_2010,schirmer_two-qubit_2009,granade_robust_2012}.
They are particularly well-suited for quantum information,
owing to their generality, robustness and the ease with which prior information can be incorporated into the algorithm.
Moveover, Bayesian approaches have been shown to allow for near-optimal Hamiltonian learning  in simple analytically tractable cases~\cite{ferrie_how_2013,granade_robust_2012}.

Bayes' theorem provides the 
proper way to re-assess, or update, prior beliefs about the Hamiltonian for a system given an experimental outcome and a distribution describing prior beliefs.  In particular,
\begin{equation}
  \label{eq:bayes-rule}
  \Pr(H | \text{data}) = \frac{\Pr(\text{data} | H)}{\Pr(\text{data})} \Pr(H),
\end{equation}
where $\Pr(H|\text{data})$ is called the posterior distribution, $\Pr(H)$ is the prior distribution that encodes our initial beliefs about
$H$ and where $\Pr( \text{data}|H)$ is the likelihood function, which
computes the probability that the observed data would occur if the Hamiltonian $H$ correctly modeled the system. The likelihood function
can be estimated by sampling from a quantum simulator for the Hamiltonian $H$ and
thus Bayesian inference causes Hamiltonian learning to reduce to a Hamiltonian simulation problem~\cite{ferrie_how_2013,granade_robust_2012,wiebe_hamiltonian_2014,wiebe_quantum_2014-1}.

Once the posterior distribution is found,
an estimate of the Hamiltonian, $\hat{H}$, is given by the expectation over the posterior,
\begin{equation}
    \label{eq:bayes-estimator}
    \hat{H} \defeq \expect[H | \text{data}]
            = \int H \Pr(H | \text{data})\ \dd H.
\end{equation}
This integral is unlikely to be analytically tractable in practice,
as it requires integrating the likelihood function $\Pr(\text{data}|H)$
over $H$.
Monte Carlo integration, on the other hand, can be much more practical.

The sequential Monte Carlo algorithm (also known as a \emph{particle filter})
provides a means of sampling from an inaccessible distribution using a transition kernel from some initial distribution \cite{doucet_sequential_2000}.
We can sample from the posterior by using Bayes' rule as the SMC transition kernel, given samples from a prior distribution and evaluations of the likelihood function.
Integrals over the posterior can then be \emph{approximated} by using these samples, which allows quantities such as $\hat{H}$ to be efficiently estimated.
SMC has seen use in a range of quantum information
tasks, including state estimation \cite{huszar_adaptive_2012},
frequency and Hamiltonian
learning \cite{granade_robust_2012},
benchmarking quantum operations \cite{granade_accelerated_2014},
and in characterizing superconducting device environments
\cite{stenberg_efficient_2014}.  Similar methods have also been applied to quantum error correction~\cite{bravyi2013simulation}.

Hamiltonians are not usually represented explicitly as matrices when using SMC algorithms and are instead parameterized by a vector $\vec{x}$
of model parameters such that $H = H(\vec{x})$.
This representation allows for parameter reduction with prior information and can include
effects outside of a purely quantum formalism, such as control distortions or
stochastic fluctuations in measurement visibility.
It also has the advantage that Hamiltonian learning is possible even in cases where matrix representations of individual terms in the Hamiltonian are not formally known.

\begin{figure*}
    \centering
    \includegraphics[width=0.8\textwidth]{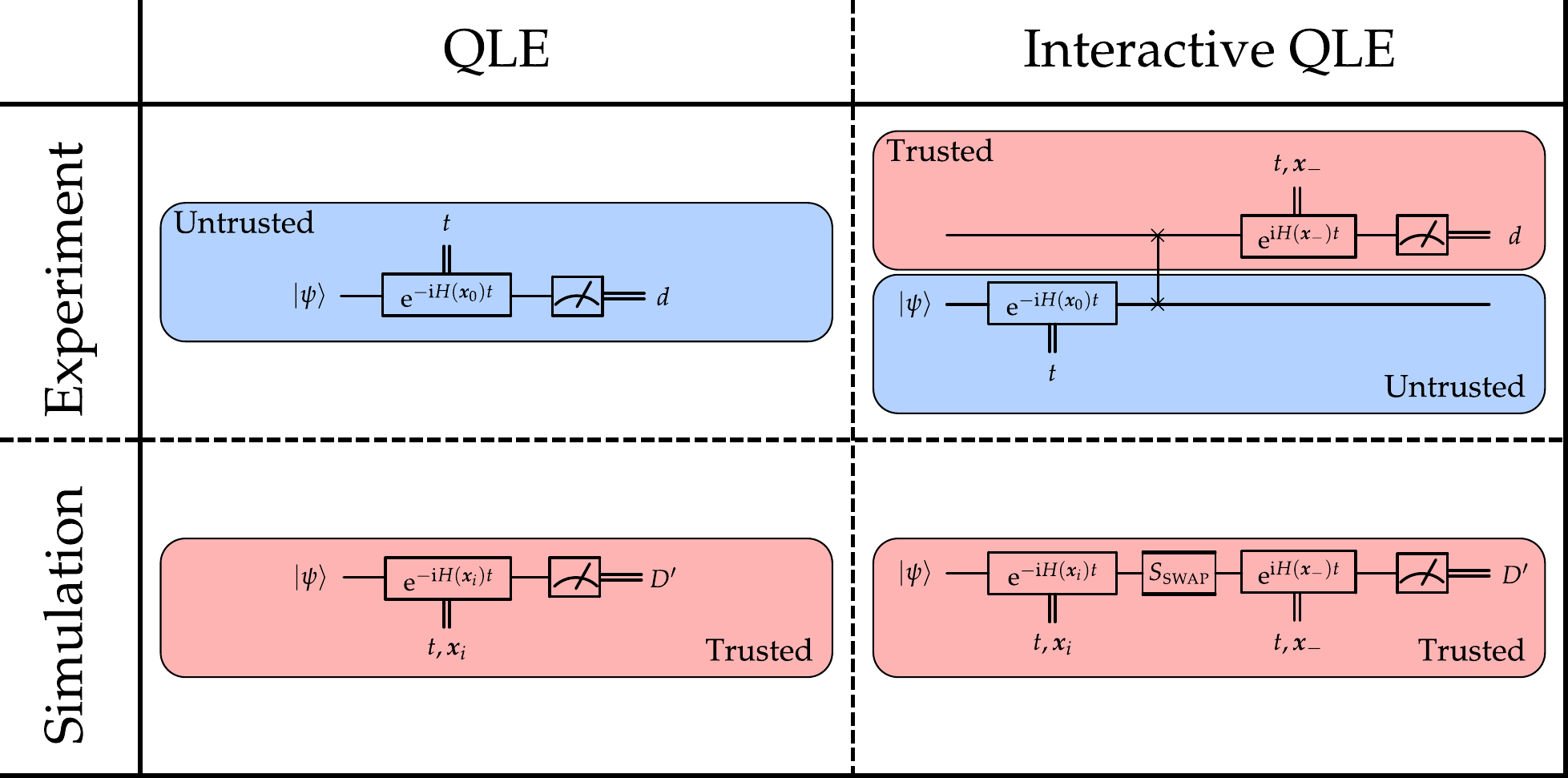}
    \caption{
        \label{fig:qhl-exp-design}
        Experiment and simulator design for (left) quantum Hamiltonian learning
        and (right) interactive quantum Hamiltonian learning with an un-truncated
        quantum simulator.  For generality, we also include $S_{\rm SWAP}$ in the simulation, which models noise or imperfections in the swap gate. Here we take $S_{\rm SWAP}=1$.
    }
\end{figure*}

Concretely, the SMC algorithm approximates prior and posterior distributions by
weighted sums of delta-functions,
\begin{equation}
    \label{eq:smc-approx}
    \Pr(\vec{x}) \approx \sum_{i=1}^{N} w_i \delta(\vec{x} - \vec{x}_i),
\end{equation}
such that the current state of knowledge can be tracked online using a
classical computer to record a list of \emph{particles}, each
corresponding to a hypothesis $\vec{x}_i$, and having a relative
weights $\{w_i\}$. These weights are then updated by substituting
the SMC approximation \eq{smc-approx} into Bayes' rule \eq{bayes-rule} to obtain
\begin{equation}
    \label{eq:smc-update}
    w_i \mapsto w_i \Pr(d | \vec{x}_i),
\end{equation}
where $d$ is an observation from the experimental system. 

Over time, the particle weights for the majority of the particles will diminish as the SMC algorithm
becomes more confident that certain hypotheses are wrong.  This reduces the total
effective number of particles in the posterior distribution and ultimately prevents learning.
This issue is addressed by using a resampling algorithm, which draws a new set of uniformly weighted SMC particles
that approximately maintain the mean and covariance matrix of the posterior distribution~\cite{liu_combined_2001}.

{Although the remainder of the process of learning using the SMC process is rigorous and well understood, error
estimates are not known for the resampling step.  Furthermore, resampling methods such as the Liu and West algorithm~\cite{liu_combined_2001}
have known pathologies where they can fail if they are provided with a multi--modal distribution or if an improbable
sequence of measurements leads to a resampling step that causes the SMC particle cloud to have no support
over the true model.  These issues are often addressed by varying the parameters used in the resampling algorithm,
majority voting on the identity of the true model over multiple runs of SMC or adjusting the guess heuristic used to choose experiments.
Moreover, these shortcomings are often heralded by effective sample size criteria built into SMC software~\cite{granade_qinfer:_2012},
such that a more appropriate resampler or set of resampling parameters can be chosen.
In spite of these theoretical shortcomings, resampling methods work exceptionally well in practice for problems in Hamiltonian learning~\cite{granade_robust_2012,wiebe_quantum_2014-1,wiebe_hamiltonian_2014,stenberg_efficient_2014}, machine learning~\cite{shan2007real}, computer vision \cite{isard_condensationconditional_1998}, and artificial intelligence~\cite{doucet_sequential_2000,arulampalam2002tutorial}.}

\subsection{Quantum Hamiltonian Learning}

Quantum Hamiltonian learning (QHL) builds upon SMC by introducing weak simulation,
in which the experimentalist has access to a ``black box'' that produces data according to an input hypothesis
$\vec{x}$. By repeatedly sampling this black box for each SMC hypothesis,
the likelihood can be inferred from the frequencies of data output by the
black-box simulator \cite{ferrie_likelihood-free_2014}. { QHL is therefore a classical Bayesian inference algorithm that uses a fast quantum method for estimating the likelihood function via quantum simulation \cite{wiebe_hamiltonian_2014}}. This augmented procedure is robust to
errors in the likelihood function introduced by finite sampling of the black box
and to approximation errors in the Hamiltonians used \cite{wiebe_quantum_2014-1}.
This latter property will be of particular importance in the development of
compressed QHL, as it allows us to use  a
\emph{truncation} of the complete system as an approximate simulator.

{The simulators used in QHL can take many forms: they could be special purpose analog simulators such as an ion trap that implements a family of transverse Ising models~\cite{korenbilt_simulation_2012}.  On the other hand, the quantum simulation could be implemented by using a quantum computer to run a digital simulation algorithm that is capable of efficiently simulating a wide array of Hamiltonian models~\cite{berry_simulation_2013} (such as $d$--sparse row--computable Hamiltonians).  We refer to all such devices as \emph{quantum simulators} to reflect the fact that they need not be a fault--tolerant quantum computer.  In our work we also require that the simulator be able to accept its initial state as input from another quantum system, but there are simpler QHL methods that do not need to be run in this fashion.}

The simplest experimental design proposed for QHL is quantum likelihood
evaluation (QLE), in which the experimenter prepares a state $\ket{\psi}$ on the untrusted
system, evolve under the ``true'' Hamiltonian $H(\vec{x}_0)$ for some time $t$, and then
measures $\{\ket{\psi}\!\!\bra{\psi}, \openone - \ket{\psi}\!\!\bra{\psi}\}$ on the trusted
simulator.  This experiment is then repeated for each SMC hypothesis
$\vec{x}_i$ until the variance in the  estimated likelihood becomes sufficiently small. 
The experiment design is illustrated in
\fig{qhl-exp-design}. QLE can be effective for learning Hamiltonians,
although it suffers from the fact that the evolution times used by the experiments must be small for most Hamiltonians.
In the case of QLE, long evolution times for typical Hamiltonians (such as Gaussian random Hamiltonians~\cite{ududec_information-theoretic_2013,gogolin_boson-sampling_2013}) produce a distribution that is very
close to uniform over measurement outcomes, such that experiments provide an exponentially small amount of information about the parameters. 

{The tendancy of quantum systems to rapidly equilibrate also causes the update step in the SMC algorithm to become unstable~\cite{wiebe_quantum_2014-1}. Here by unstable we  mean that small perturbations in the estimated likelihoods result in large deviations in the posterior distribution.  This can be combatted by using short experiments, which lead to uncertainties about the true model (after one update) that scale at least as $O(t^{-1})$ (see~\app{fisher}). As a result, short-time evolutions necessitate processing exponentially more data than would be required if long experiments could be performed. Moreover, given that the time required for state preparation is independent of $t$, it is clear that the total experimental time used to learn the true model will be prohibitively large in such cases.  Thus the ability to use long experiments can lead to substantial improvements for Hamiltonian learning}.

To use the long evolution times requisite
for expedient high-accuracy characterization the system of interest can be coupled to the
simulator using $\swapgt$ gates, as shown in
\fig{qhl-exp-design}. This experiment design, interactive quantum likelihood
evaluation (IQLE), uses the simulator to approximately \emph{invert} the forward
evolution under the unknown system, such that the measurement is approximately
described by $H(\vec{x}_0) - H(\vec{x}_-)$ provided we choose an inversion
hypothesis $H(\vec{x}_-)$ that
is close to $H(\vec{x}_0)$. Intuitively, such experiments directly compare the dynamics of $H(\vec{x})$ and $H(\vec{x}_-)$.  Such experiments also reduce the norm of the effective system Hamiltonian, which typically allows the system to evolve for much longer before 
  the quantum probability distribution
becomes flat.  {These SWAP gates need not be perfect, as the learning protocol is known to be robust to such errors~\cite{wiebe_quantum_2014-1}.  We further discuss the effects of faulty SWAP gates on bootstrapping in~\app{booterror}.
In cases where a SWAP gate cannot be implemented, such as when the trusted resource is implemented using an
incompatible modality, non-interactive quantum Hamiltonian learning can be used to perform the first
iteration of bootstrapping, such that SWAP gates are available in all further iterations.
That is, we can use QLE to initialize the bootstrapping procedure, and can proceed
using IQLE.}

In order to combat the exponentially diminishing likelihood of the system
returning to its initial state after the inversion, we require that
$\|H(\vec{x}) - H(\vec{x}_-)\| t$ is approximately
constant~\cite{wiebe_hamiltonian_2014}.  We use the particle guess heuristic
(PGH) to achieve this.  The PGH involves drawing two hypotheses about $H$,
$\vec{x}_-$ and $\vec{x}_-'$, from the prior distribution and then choosing $t
= 1 / \|H(\vec{x}_-) - H(\vec{x}_-')\|$. Since $\|H(\vec{x}_-) -
H(\vec{x}_-')\|$ is an estimate of the uncertainty in the Hamiltonian, we
expect that at most a constant fraction of the prior distribution will satisfy
$\|H(\vec{x}_-) - H(\vec{x})\|t \ge 1$ assuming $H(\vec{x})$ is linear and the
prior distribution has converged to a unimodal distribution centered about the
true Hamiltonian.  The heuristic therefore seldom leads to experiments for
which $|\bra{\psi}e^{iH(\vec{x}_-)t}e^{-iH(\vec{x})t}\ket{\psi}|^2\sim 1/2^n$
for most $\vec{x}$. Moreover, since the PGH relies only on the current SMC
approximation to the posterior, the heuristic incurs no additional simulation
costs. Rather, the PGH provides adaptivity by depending on the current state
of knowledge about the state of the quantum system through the particles
$\vec{x}_-$ and $\vec{x}_-'$. Experiment design via the particle guess
heuristic has been shown to lead to efficient estimation of Hamiltonians
using IQLE \cite{wiebe_hamiltonian_2014}, and has since been usefully applied
in other experimental contexts \cite{stenberg_efficient_2014}.

Previous work has analyzed the complexity of learning using IQLE~\cite{wiebe_hamiltonian_2014}.
In cases where the error in the characterized Hamiltonian scales as $e^{-\gamma N_{\exp}}$ and $N_{\rm samp}$ samples are used to 
estimate the likelihood function, the protocol requires $O(N_{\rm samp} N_{\rm particles}\log(1/\epsilon)/\gamma)$ simulations to learn
the vector of Hamiltonian parameters to within error $\epsilon$ as measured by the $2$--norm.  In practice, the decay constant $\gamma$ depends on the number of parameters used to describe $H$ and the properties of the experiments used.  It does not directly depend on the dimension of $H$~\cite{wiebe_hamiltonian_2014,wiebe_quantum_2014-1}.  The updating procedure used to combine these results is further known to be stable provided that
the likelihoods of the observed experimental outcomes are not exponentially small for the majority of the SMC particles~\cite{wiebe_hamiltonian_2014}.  This occurs for
well posed learning problems that use two outcome experiments.

\section{Compressed QHL}\label{sec:QHL}

\begin{figure}
    \includegraphics[width=0.95\linewidth]{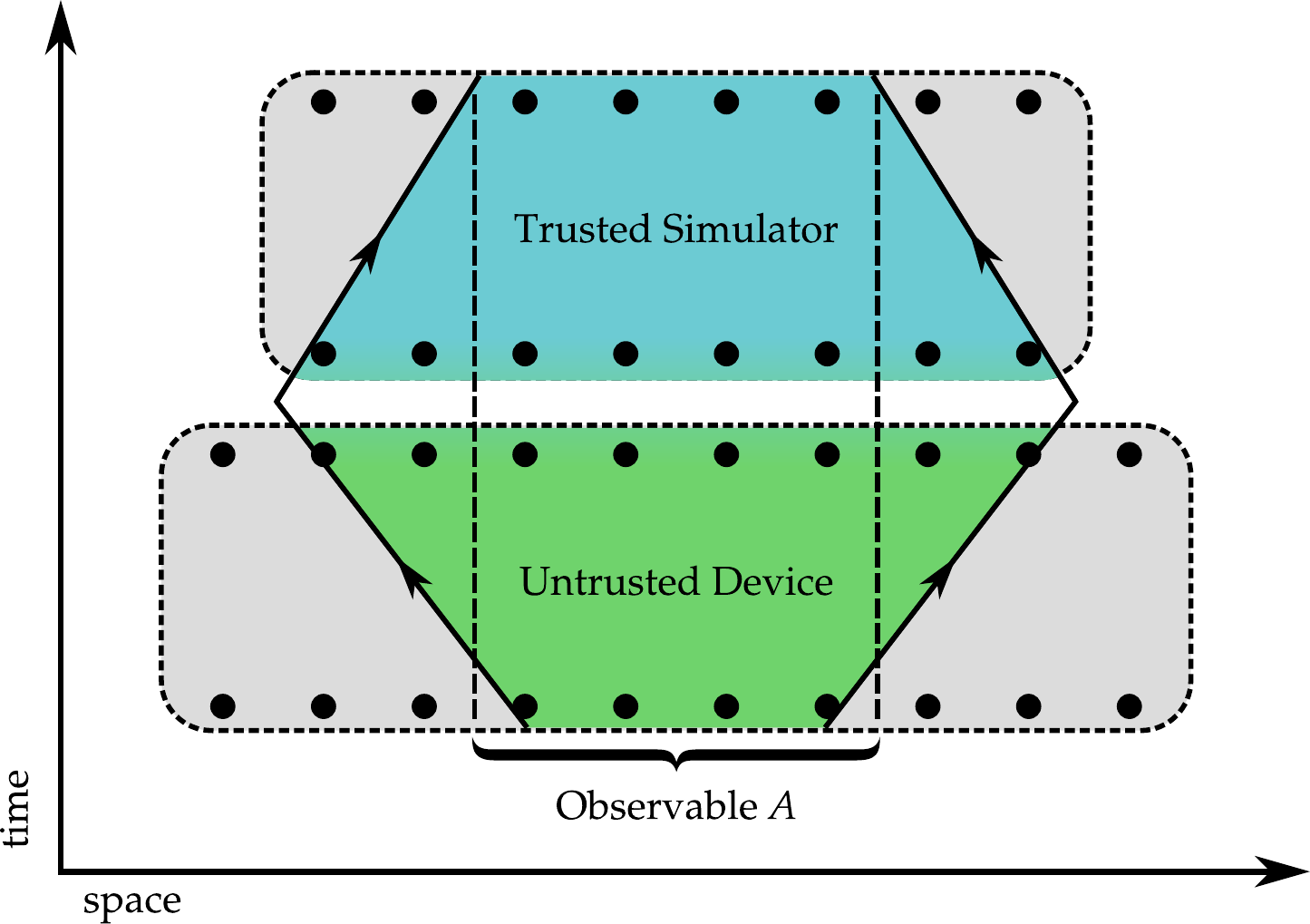}
    \caption{
        Light cones for $A(t)$ for a single step of an $r$ step protocol.
        The green region is the light cone after the evolution in the untrusted device,
        and the blue region is after inversion in the trusted device.
        The dashed lines show the spread of $A(t)$ due to inexact inversion in the trusted simulator.
    }
    \label{fig:lightconeb}
\end{figure}

Information locality is what enables cQHL and in turn quantum bootstrapping.
This idea is made concrete via Lieb--Robinson bounds, which show that an analog of special relativity exists for local observables evolving under Hamiltonians that have rapidly decaying interactions~\cite{hastings_spectral_2006,nachtergaele_lieb-robinson_2006,hastings2010locality,da_silva_practical_2011}.
Lieb--Robinson bounds give an effective ``light cone'', as illustrated in \fig{lightconeb},
in which the evolution of an observable $A$ can be accurately simulated without needing to consider any subsystem outside of the light cone.
Specifically, they imply that a local observable $A(t)$ provides
at most an exponentially small amount of information about subsystems that are further than distance $st$ away from the support of $A(0)\equiv A$, where $s$ is the Lieb--Robinson velocity for the system and $t$ is the evolution time.
Here, $s$ is analogous to the speed of light, and only depends on the geometry and strengths of the interactions in the system~\cite{hastings_spectral_2006,nachtergaele_lieb-robinson_2006,hastings2010locality}.  Thus, if $st$ is bounded above by a constant and the initial support of $A$ is small then the support of $A(t)$ is at most a constant.  This shows that the dynamics of $A(t)$ can be simulated using a small quantum device, provided $st$ is sufficiently small.

Compressed QHL exploits this intuition by evolving an initial state under the untrusted quantum
simulator, swapping the quantum state of a subsystem of the larger
(uncharacterized) system into a quantum simulator, and then approximately
inverting the evolution by guessing the Hamiltonian dynamics and simulating
their inverse.  It then measures the simulator to determine whether the
inversion yielded the initial state or a state in its orthogonal compliment.
One step of this process is illustrated in~\fig{lightconeb}.

The inversion process used in interactive QLE not only leads to more
informative experiments, but we will show in \sec{noncomm} that generalizing to include
repeated applications of swaps and inverse simulations also delay the rate at
which the light cone propagates from the observable. This in turn allows much
longer evolution times to be used without the observable stretching beyond the
confines of the trusted simulator.

In particular, this swapping procedure leads to characteristic Lieb--Robinson
velocities that shrink as the experimentalist learns more about the system.
\emph{That is, the light cone represents an ``epistemic'' speed of light in
the coupled systems that arises from the speed of information propagation
depending more strongly on the uncertainty in the Hamiltonian than the
Hamiltonian itself.} Since the effective speed of light slows as more
information is learned, long evolutions can be used when the uncertainty is
small. This removes a major restriction of the method of Da Silva et
al~\cite{da_silva_practical_2011} since the variance of any unbiased estimator of the
Hamiltonian parameters diverges as $\Omega(t^{-2})$ for Hamiltonian learning methods that
use short time experiments (see~\app{fisher} for more details).

\begin{figure}
    \centering
    \includegraphics[width=0.7\columnwidth]{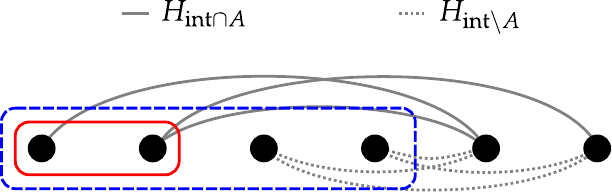}
    \caption{
        \label{fig:bootstrapping-partition}
        Separation of $\Hint=\Hintcap+\Hintminus$ where 
        $\Hintcap$ are interactions with qubits in the support of $A$ (red solid box) and $\Hintminus$ interacts
with qubits that are swapped into the trusted simulator but are outside $A$ (blue dashed box).
    }
\end{figure}

To analyse the error introduced by compressed simulation, we consider learning
a Hamiltonian $H$ by measurement of an observable $A$ supported on $a$ sites,
using a simulator with support on $w > a$ sites. We then expand $H$
into those terms $\Hin$ which we can access with our simulator, the terms
$\Hout$ supported only outside of the simulator entirely, and the interaction
$H_{\text{int}}$ between these two partitions. {We then further break down
$H_{\text{int}}$ into those terms $\Hintcap$ which have non--trivial action between sites in $A$
and the neglected sites, and those terms $\Hintminus$ which act upon sites in
the simulator, but not the observable}. That is, we expand $H$ as
\begin{gather}
  \label{eq:bs-comm-Hexpand}
  \begin{aligned}
    H & = H_{{\rm int}} + H_{\rm in} +H_{\rm out} \\
      & = H_{{\rm int}\bigcap A}+H_{{\rm int}\setminus A} + H_{\rm in} +H_{\rm out}.
  \end{aligned}
\end{gather}
The decomposition of the interaction Hamiltonian $H_{\mathrm{int}}$ into
couplings that include and exclude $A$ is illustrated in
\fig{bootstrapping-partition}.

cQHL neglects the
terms included in $\Hintcap$ when processing data collected from the system
of interest. If $H$ exhibits a finite Lieb-Robinson velocity, then we can
bound the error introduced by this approximation. Moreover, we will show
that by using interactivity as a resource, we can reduce this error as
we become more certain about the dynamics of the untrusted system.
We discuss these two considerations in more detail below.

\subsection{Commuting Hamiltonians}

It is helpful, however, to first build intuition by considering the special
case that all terms in the unknown system's Hamiltonian are local and mutually
commute. This is true, for instance, in the Ising models~\eq{ising-model} that
we consider in numeric examples. In this case, the compressed interactive
likelihood evaluation experiment described in \fig{qhl-exp-design} is
particularly simple to analyze.

If we work in the Heisenberg picture then it is easy to see from the
assumption that the Hamiltonian terms commute with each other (but not necessarily $A$) that $[H_{{\rm int}\setminus A} + H_{\rm out}, A(t)]=0 $.  This implies that
\begin{align}
A(t)&=e^{i H_{\rm in} t}e^{i H_{{\rm int}\bigcap A} t}Ae^{-i H_{\rm in} t}e^{-i H_{{\rm int}\bigcap A} t}\nonumber\\
\tilde{A}(t)&=e^{i H_{\rm in} t}Ae^{-i H_{\rm in} t},
\end{align}
where $\tilde{A}(t)$ is the simulated observable within the trusted simulator. 

Using Hadamard's Lemma and the triangle inequality to bound the
truncation error $\|\tilde{A}(t) - A(t)\|$,
we obtain that
\begin{align}
\|\tilde{A}(t) - A(t)\|&\le \|A\|(e^{2\|H_{{\rm int}\bigcap A}\| t}-1)\label{eq:comm}
\end{align}
If the objective is to have error at most $\delta$ in the compressed simulation then it suffices to choose experiments with evolution time at most
\begin{equation}
t\le \ln\left( \frac{\delta}{\|A\|}+1\right)\left(2\|H_{{\rm int}\bigcap A}\| \right)^{-1}.\label{eq:tbd}
\end{equation}
If the sum of the magnitudes of the interaction terms that are neglected in the simulation is a constant then~\eq{tbd} shows that $t$ scales at most linearly in $\delta$ as $\delta\rightarrow 0$.  This is potentially problematic because short experiments can provide much less information than longer experiments so it may be desirable to increase the size of the trusted simulator as $\delta$ shrinks to reduce the experimental time needed to bootstrap the system.  QHL is robust to $\delta$~\cite{wiebe_hamiltonian_2014,wiebe_quantum_2014-1} and $\delta\approx 0.01$ often suffices for the inference procedure to proceed without noticeable degradation.

Note that if $H_{{\rm int}\bigcap A} =0$ then infinite--time simulations are possible for commuting models (such as Ising models) because no truncation error is incurred.  Non--trivial cases for QHL only occur in commuting models with long range interactions.

\subsection{Non--commuting Hamiltonians}
\label{sec:noncomm}

If the Hamiltonian contains non--commuting terms then the factorization of $e^{-iHt}$ used in~\eq{comm} no longer holds.  This is because 
\begin{gather}
  \begin{aligned}
  &e^{i(H_{\rm in} + H_{\rm int} + H_{\rm out})}Ae^{-i(H_{\rm in} + H_{\rm int} + H_{\rm out})} \\
  &\qquad \ne e^{i(H_{\rm in} + H_{\rm int})}Ae^{-i(H_{\rm in} + H_{\rm int})},
\end{aligned}
\end{gather}
unlike in commuting models.
Such dynamics can also lead to observables $A(t)$ that rapibly obtain non--negligible support near the boundary of the trusted simulator.  The trusted system will not tend to simulate these evolutions accurately because significant interactions exist between $A(t)$ and the neglected portion of the system.  This means a more careful argument will be needed to show that bootstrapping will also be successful here.

We address this issue by generalizing IQLE experiments.  Typically, each IQLE
experiment is of the form $e^{iH_- t}Se^{-iHt}$, where $S$ is a swap operation
and $H_-$ is a Hamiltonian simulated by the trusted simulator.  Instead of
swapping the states of both devices once, we generalize such experiments to
consist of $r$ swaps, such that the observable evolves under the unitary
$(e^{iH_-t/r}Se^{-iHt/r})^r$ before we perform a measurement.  It is then easy
to see from the Trotter formula that $e^{iH_-t/r}e^{-iHt/r}\approx
e^{-i(H-H_-)t/r}$. The inclusion of additional swap operations serves two
purposes.  Firstly, it causes the terms in the Hamiltonian to effectively
commute with each other for $r$ sufficiently large.  Secondly, if $t$ is small
relative to $\|H-H_-\|^{-1}$ then $r$ swaps of the two registers will not
cause the $A(t)$ to have substantial support on the boundary of the trusted
simulator at any step in the protocol.

If a large value of $r$ is chosen, then the system effectively evolves under
$e^{-i(H-H_-)t}=e^{-i(H_{\rm out} + H_{\rm int} + \Lambda)t}$, where
$\Lambda:=H_{\rm in} - H_-$.  We expect that the dynamics of $A$ will
therefore be dictated by the properties of $\Lambda$ for short evolutions.  We
make this intuition precise by showing in \app{lrb} that the
error from using a small trusted simulator obeys
\begin{widetext}
  \begin{gather}
    \begin{aligned}
    \label{eq:lrberror}
      \|A(t) - \tilde{A}(t)\| &\le (\|[H_{\rm in},\Lambda]\| + \|[H_{\rm int},H_{\rm in}]\|)\|A\|\frac{t^2}{r}+2\|H_{{\rm int}\bigcap A}\|\|A\| t\\
      &\qquad+2\|H_{{\rm int}\setminus A}\| \|A\| |\{A\}|te^{-\mu {\rm dist}(A,H_{\rm out})}\left[e^{2s|t|} -1 \right]e^{2\|H_{\rm out}+H_{{\rm int}\setminus A}\|t/r},
    \end{aligned}
  \end{gather}
\end{widetext}
for cases of nearest--neighbor or exponentially decaying interactions between subsystems.  {$H_{\rm in}$, $H_{\rm int}$, $H_{\rm out}$ and related terms are explained in~\eq{bs-comm-Hexpand} and the surrounding text.}
Here $s$ is the Lieb--Robinson velocity for evolutions under $\Lambda$ and $\mu$ is related to the rate at which interactions decay with the graph distance between subsystems.
 It is worth noting that~\eq{lrberror} can be improved by using higher order Trotter--Suzuki formulas in place of the basic Trotter formula to reduce $r$~\cite{suzuki_fractal_1990} and  also by using tighter Lieb--Robinson bounds for cases with nearest--neighbor Hamiltonians.

The variable $\Lambda$ is related through the particle guess heuristic to the uncertainty in the Hamiltonian, which implies that the speed of information propagation is also a function of the uncertainty in $H$~\cite{hastings_spectral_2006,nachtergaele_lieb-robinson_2006}.  That is,  longer evolutions can be taken as $H$ becomes known with ever greater certainty.  This means that the  Lieb--Robinson velocity does not pose a fundamental restriction on the evolution times permitted because $s\rightarrow 0$ as $\Lambda \rightarrow 0$.  

Of course, the error term $2\|\Hintcap\|t$ in~\eq{lrberror} places a limitation on the evolution time but that term
can be suppressed exponentially by increasing the diameter of the set of qubits in the trusted simulator for systems with interactions that decay at least exponentially with distance.  Thus the roadblocks facing compressed QHL can be addressed at modest cost by using our strategy of repeatedly swapping the subsystems in the trusted and untrusted devices.

As an example, if we assume (a) that the interactions are between qubits on a line (b) that $w-a$ is chosen such that $8st/\mu<{w-a}$ then in the limit as $r\rightarrow \infty$
\begin{equation}
t\le \frac{\delta}{2\|A\|(\|H_{{\rm int}\bigcap A}\|\!+\! 2\| H_{{\rm int}\setminus A}\| |\{A\}|e^{-\mu (w-a)/4})} \label{eq:lrberror2},
\end{equation}
suffices to guarantee simulation error of $\delta$.  This result is qualitatively similar to~\eq{tbd} and requires that $|w-a|$ scales at most logarithmically with the total evolution time desired.


\subsection{Scanning}

\begin{figure}
  \includegraphics[width=0.8\linewidth]{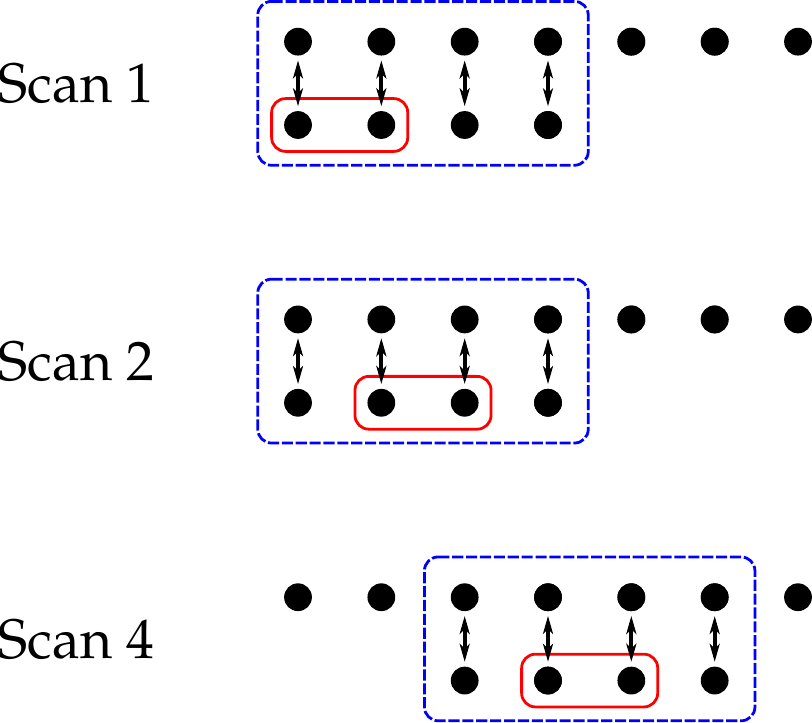}
  \caption{Scanning procedure for $7$ qubits, a $4$ qubit simulator and a $2$ qubit observable.  Blue (dashed) box is support of simulator, red (solid) box is support of $A$.}
  \label{fig:scan}
\end{figure}

The previous methods provide a method for characterizing a subsystem of the global Hamiltonian.  These results cannot be used
directly to learn the full system Hamiltonian because the trusted simulator lacks couplings present in the full system.  Instead the Hamiltonian 
must be inferred by patching together the results of many Hamiltonian inference steps.  This process can be thought of as a scanning procedure wherein
an observable is moved across the set of qubits collecting information about the couplings that strongly influence it.  The scanning procedure is illustrated in~\fig{scan}.

In order to properly update the information about the system, we modify
sequential Monte Carlo to use two particle clouds. The first is the global
cloud, which keeps track of the prior distribution over all the parameters in
the Hamiltonian model.  The second is the local cloud, which keeps track of
all of the parameters needed for the current compressed QHL experiment. The
global cloud is constrained such that the weights of each particle in the
cloud is constant (i.e. the probability density is represented by the density
of particles rather than their weight), whereas the local cloud is not
constrained in this fashion. This constraint on the global cloud is needed
because resampling does not in general preserve the indices of each particle,
so that there is no way to sensibly identify a global particle that
corresponds to a particle in the local posterior.

Instead, by copying a subset of global parameters into the local cloud, our
modified particle filter approximates the prior by a product distribution
between the local and remaining parameters. Resampling the local posterior
then makes this approximation again, ensuring that the local weights are
uniform. Thus, we can copy the (newly resampled) local cloud into the global
cloud, overwriting the corresponding parameters. Once the local cloud is
merged back into the global cloud in this way, we begin the next step in the
scan by selecting a different set of parameters for the local cloud, and
continuing with the next compressed QHL experiment.

We implement this scanning procedure in our numerical experiments by using
a local observable centered as far left on the spin chain as possible.  We then infer the Hamiltonian for this location using a fixed number of
experiments, swap the Hamiltonian parameters from each of the SMC particles to the global cloud and then move the observable one site to the right.  This process is repeated until the observable has scanned over the entire chain of qubits, and then we begin again by scanning over the first $2a$ qubits in reverse, where $a$ is the width of the observable.  We do this to reduce the systematic bias that emerges from the fact that Hamiltonian parameters associated with couplings learned earlier in the procedure will have greater uncertainty.

\subsection{Complexity of cQHL}
\label{sec:cqhl-complexity}

Here 
investigate the circumstances under which cQHL is efficient.  The
number of experiments that are needed in cQHL is $N_{\rm exp} = N_{\rm scans}
N_{\rm exp/scan}$.  If we then make the pessimistic assumption that $O(n)$
scans are needed and assume that the error within each scan decays as
$e^{-\gamma N_{\rm exp/scan}}$ then the number of experiments needed to make
the \emph{combined error} in the inferred vector of Hamiltonian parameters, $\vec{x}$, at most $\epsilon$ it suffices to use a number of experiments that scales as
\begin{equation}
N_{\rm exp} \in O\left(\frac{n\log(n/\epsilon)}{\gamma} \right).
\end{equation}
Here we have used the fact that if there are $n$ scans and the error in the Hamiltonian parameters is $\epsilon/n$ then the error in the reconstructed Hamiltonian parameters after $n$ scans is at most $n\times\epsilon/n=\epsilon$.  This bound is again pessimistic as in practice the information learned in subsequent scans actually reduces, rather than increases, the error in parameters learned from previous scans.

The number of calls to our trusted simulator needed to update the weights of
the particles in the SMC cloud is $N_{\rm sim} = N_{\rm exp}N_{\rm samp}N_{\rm part}$. Firstly, if $N_{\rm samp}\in O(1/\delta^2)$ samples are
used to infer the likelihoods of the outcome for each of the $N_{\rm part}$
particles in the SMC cloud, then the total number of experiments required to
learn a fixed Hamiltonian on $n$ qubits the number of simulations that are
needed to estimate the likelihoods within error $\delta$ is
\begin{equation}
N_{\rm sim} \in O\left(N_{\rm part} \times \frac{n \log(n/\epsilon)}{\gamma\delta^2} \right).
\end{equation}
This quantifies the number of experiments that are needed to estimate the likelihoods within error $O(\delta)$ given a perfect simulator.  

In cases with non-commuting Hamiltonians, error is also incurred by using a non-infinite value of $r$.
If we also demand that the contribution from this source of error is $O(\delta)$ then it is straight forward to see from~\eq{lrberror} that the necessary value of $r$ scales at most as
\begin{widetext}
\begin{equation}
r\in O\left(\frac{(\|[H_{\rm in},\Lambda]\|+\|[H_{\rm int},H_{\rm in}]\|)\|A\|t^2}{\delta} + (\|H_{\rm out}\|+\|H_{\rm in\setminus A}\|)t \right).
\end{equation}

The above relations set the complexity (as measured by the number of
experiments and the number of swaps) of Hamiltonian characterization, but the
space requirements also are problem dependent in cases that have non-commuting
Hamiltonians. If we assume that all interactions between arbitrary qubits $x$
and $y$ decay at least as $e^{-\nu {\rm dist}(x,y)}$, then the distance between
$A$ and the neglected part of the Hamiltonian $\Hintminus$ can be chosen as
\begin{equation}
{\rm dist}(A,H_{\rm out}) \in O\left(\frac{1}{\min\{\mu,\nu\}}\left(st+ \log\left(\frac{\|H_{{\rm int}\setminus A}\|\|A\||\{A\}|t}{\delta} \right) \right) \right),
\end{equation}
\end{widetext}
where $|\{A\}|$ is the number of sites on which $A$ is supported, $s$ is the
Lieb-Robinson velocity for evolution under $\Lambda$, and where $\mu$ is the
exponential clustering parameter used in \eq{lrberror}.

From the particle guess heuristic, we have that $st$ is asymptotically
constant and thus both the number of qubits and the number of experiments
scale logarithmically with the desired accuracy.  The number of inversions
used $r$ scales polynomially with $t$, which scales as
$O(1/\epsilon)$~\cite{wiebe_hamiltonian_2014} and thus the number of swaps
scales polynomially with the desired error tolerance.  This scaling can be
made to approach the Heisenberg limit of $O(1/\epsilon)$ scaling by replacing
the Trotter formula used in the inversion step with increasingly high--order
Trotter--Suzuki formulas as $\epsilon$ shrinks~\cite{suzuki_fractal_1990}.

The above analysis can also be extended to systems with polynomial decay of
interactions. In such cases, the cQHL algorithm is less efficient because the
logarithmic scaling with $1/\epsilon$ is replaced with polynomial scaling with
$1/\epsilon$ in most instances of polynomially-decaying interactions.
Even in such cases, however, we expect the sequential Monte Carlo algorithm
to remain robust to simulation errors such as truncation. Thus, we expect that
compressed quantum Hamiltonian learning can offer advantages when the
Lieb-Robinson velocity of an untrusted system under inversion by a hypothesis
is finite.

In some cases, such as dipolar coupling, existing Lieb--Robinson bounds
diverge~\cite{hastings_spectral_2006}. This does not imply, however, the lack
of a finite information propagation velocity however; indeed, it is worth
noting that finite speeds of information propagation are expected
theoretically and observed experimentally for finite systems with dipolar
couplings~\cite{richerme_non-local_2014}. Moreover, in the presence of disorder,
experimental evidence suggests that the information propagation velocity
can be dramatically reduced~\cite{alvarez_quantum_2013}.

An important remaining issue is the scaling of $\delta$ and the number of particles.  We know from prior work that Bayesian inference is highly tolerant of errors in the likelihood function~\cite{wiebe_quantum_2014-1} and that $\delta\approx 0.01$ typically suffices.  Furthermore, The number of particles in the SMC cloud, $N_{\rm part}$, is a slowly increasing function of the number of model parameters for the
Hamiltonian and does not explicitly depend on the Hilbert space dimension~\cite{beskos_stability_2011}.  In the numerical experiments we have performed in this, and prior work, we observe that $N_{\rm part}$ tends to scale roughly logarithmically with the number of model parameters~\cite{granade_robust_2012, wiebe_quantum_2014-1,wiebe_hamiltonian_2014}.  This means that none of the above issues present a fundamental obstacle for compressed quantum Hamiltonian learning.

\section{Quantum Bootstrapping}\label{sec:bootstrapping}

We will now turn our attention to quantum bootstrapping, which is an
application wherein compressed quantum Hamiltonian learning is used to infer
control maps for uncharacterized devices. Control maps relate control settings
of a device to its system Hamiltonian.  Learning these maps is of particular
importance if cross-talk or defects cause different parts of the system to
respond differently to the same controls. In such cases, Hamiltonian
characterization is a necessary part of the control design and calibration
process.

To concretely show how quantum Hamiltonian learning can address this challenge, we
consider a model in which a row--vector of control settings $\vec{C}$ is related to the system Hamiltonian
by an affine map $H(\vec{C})$,
\begin{equation}
    H(\vec{C}) = \vec{C} \cdot [H_1,\ldots, H_M] + H_0\label{eq:Hvec}
\end{equation}
for some $M + 1$ unknown Hamiltonians $\{H_0, \ldots, H_M\}$. By the same
argument as before, let $H_j = H(\vec{x}_j)$ be represented by a model parameter
vector, such that this is an efficient representation of the control landscape.

The control learning process then proceeds as follows:
\begin{itemize}
  \item[(a)] Set $\vec{C}=\vec{0}$ and learn $H_0$ using compressed QHL.
  \item[(b)] For $k=1,\ldots, M$ set $C_j=\delta_{j,k}$  and learn $H_k+H_0$
      using compressed quantum Hamiltonian learning.
  \item[(c)] Subtract $H_0$ from these values to learn the vector $\vec{v}_k$ that describes the model for $H_k$.
\end{itemize}
This yields a vector of Hamiltonian parameters that describes each control term $H_k$.  If we then imagine the matrix $G$ such that $G_{k,j}=[\vec{v}_k]_j$ then a model for $H(\vec{C})$ is given by $G \vec{C}^T$, which allows the effect of control on the quantum system of interest to be predicted.

Non--linear controls can be
learned in a similar fashion by locally approximating the control function with a piecewise--linear function.

We complete our description of quantum bootstrapping by detailing how control
learning can be used to calibrate an initially untrusted device.  
If $H(\vec C)$ is an affine map then this can be accomplished using the following approach:
\begin{itemize}
  \item[(a)] Learn $H(\vec{C})$ using the above method.
  \item[(b)] Choose a set of fundamental Hamiltonian terms, $\mathcal{H}_j$,
    from which all Hamiltonians in the class of interest can then be generated.
  \item[(c)] For each $\mathcal{H}_j$ apply the Moore--Penrose pseudoinverse
    of $G$ to $\mathcal{H}_j -H_0$ to find $\vec{C}_j$ such $H(\vec{C}_j) \approx \mathcal{H}_j$.
  \item[(d)] Treat the system as a trusted simulator and repeat steps (a), (b) and (c) for a larger system.  
\end{itemize}

In cases where $H_0=0$, $H(\vec{C})$ is linear and hence
$H(a\vec{C}_1+b\vec{C}_2)= a\mathcal{H}_1+b\mathcal{H}_2$. 
This means that an arbitrary Hamiltonian formed from a linear combination of the
$\mathcal{H}_j$ can be implemented.  If $H_0\ne 0$, then this process is less straightforward. 
It can be solved by applying a pseudoinverse to find a control vector $\vec{C}(a, b)$ that produces
$a\mathcal{H}_1 +b\mathcal{H}_2$, but such controls will be specific to $a$ and $b$. 
A simple way to construct a general control sequence is to use Trotter--Suzuki formulas to approximate the dynamics
in terms of $\mathcal{H}_1 = H(\vec{C}_1)$ and $\mathcal{H}_2 = H(\vec{C}_2)$ as
\begin{equation}
\left(e^{-i\mathcal{H}_1a\Delta t / R}e^{-i\mathcal{H}_2b\Delta t/R}\right)^R\!=\! e^{-i(a\mathcal{H}_1+b\mathcal{H}_2)\Delta t} + O\left(\frac{\Delta t^2}{R}\right).
\end{equation}
Higher--order Trotter--Suzuki methods can be used to reduce the value of $R$ if desired~\cite{suzuki_fractal_1990}.

Errors accrue as the bootstrapping procedure progresses. However, since the
error shrinks exponentially with the number of experiments for well-posed
learning problems, the number of experiments needed per recursion will often
scale linearly with the total number of intermediate untrusted devices needed
to reach the final system of interest (see \app{booterror}).

\begin{figure}
\includegraphics[width=0.75\linewidth]{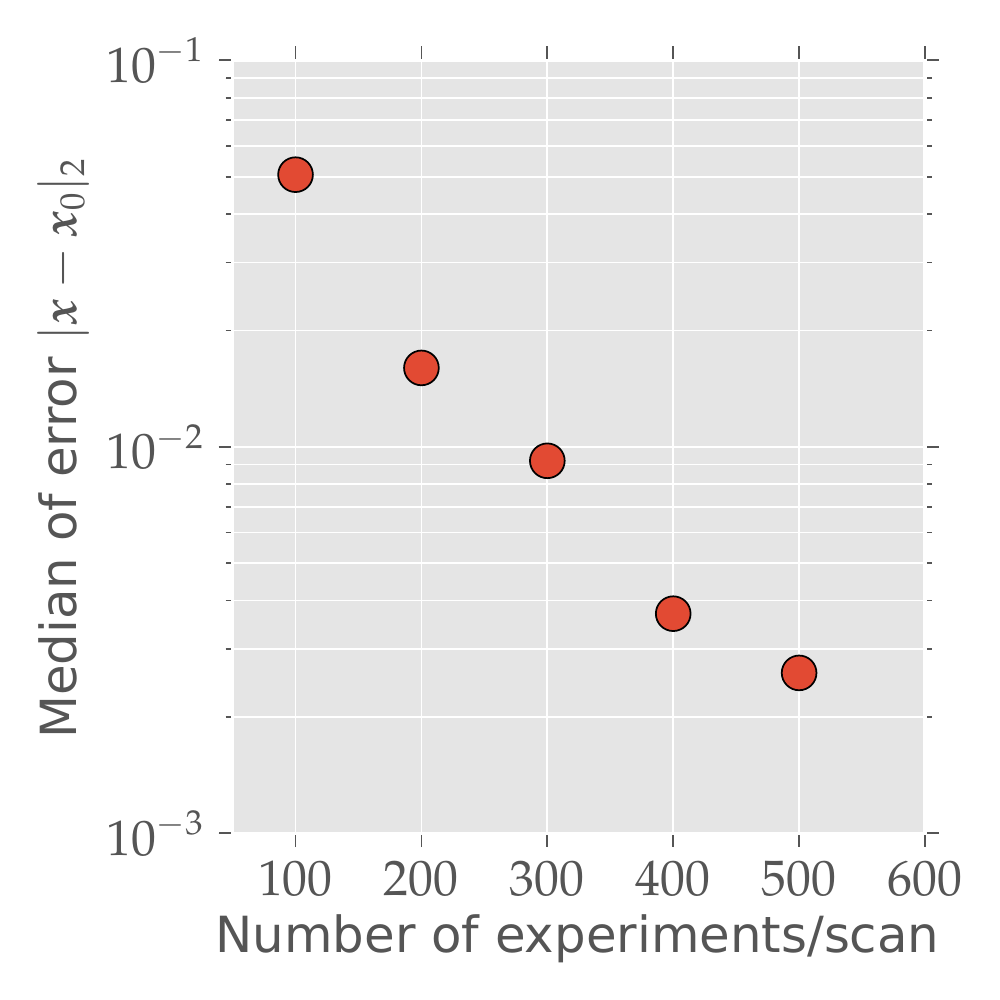}
\caption{Error in QHL for $a=4$ with varying $N_{\exp}$ per scan.  Data consistent with $e^{-0.006 N_{\exp}}$ scaling.}\label{fig:experr}
\end{figure}


\subsection{Error Propagation in Bootstrapping}
\label{app:error-bounds}

Let $G$ be the control map that is inferred via the inversion method discussed above and let $G+\mathcal{E}$ be the actual control map that the system performs.
If we measure the error in a single step of bootstrapping to be the operator norm of the difference between the bootstrapped and the target Hamiltonians then we have that the control error for the system after bootstrapping obeys
\begin{equation}
\|(G+\mathcal{E})G^+ \mathcal{H}_k - \mathcal{H}_k\|\le (\|GG^+ - \openone\| + \|\mathcal{E}\|\|G^+\|)\|\mathcal{H}_k\|.\label{eq:Gamma}
\end{equation}
where $G^+$ is the pseudo--inverse of $G$.
Eq.~\eq{Gamma} shows that the error after a single step is a multiple of the norm of the control Hamiltonian that depends not only on the error in the compressed QHL algorithm but also on $\|GG^+ - \openone\|$ which measures the invertibility of the control map.  Since the error is a multiplicative factor, it should not come as a surprise
that the error after $L$ bootstrapping steps grows at worst exponentially with $L$.  In particular, the bootstrapping error is at most
\begin{equation}
L\Gamma_{\max} e^{(L-1) (\kappa_{\max}-1+\|\mathcal{E}_{\max}\|\|G^+_{\max}\|)}\max_k \|H_k\|,\label{eq:errscale}
\end{equation}
where $\Gamma_{\max}$ is the maximum value of $ (\|GG^+ - \openone\| + \|\mathcal{E}\|\|G^+\|)$ over all the $L$ bootstrapping steps, $\kappa_{\max}$ is the maximum condition number for $G$, $\|\mathcal{E}_{\max}\|$ and $\|G^+_{\max}\|$ are the maximum values for the error operator and the pseudoinverse of $G$ over all $L$ steps.  The proof of~\eq{errscale} is a straight forward application of the triangle inequality and is provided in \app{booterror}.

Given that the error tolerance in the bootstrapping procedure is $\Delta\le 1$, $G$ is invertible and that $w$, $a$ and $t$ are chosen such that $\|\mathcal{E}_{\max}\|\le e^{-\gamma N_{\exp}}$ (i.e. a constant fraction of a bit is learned per experiment) it is easy to see that~\eq{errscale} is less than $\Delta$ if
\begin{equation}
N_{\exp} \ge \frac{(L-1)\kappa_{\rm max}+\ln\left(\frac{L\|G_{\max}^+\|\max(\max_k\|H_k\|,1)}{\Delta}\right)}{\gamma}.\label{eq:efficiency}
\end{equation}
This process is efficient provided $\gamma$ is at most polynomially small.  If $G$ is not invertible, then
the error cannot generally be made less than $\Delta$ for all $\Delta>0$.
Although, if $G$ is not invertible then the system is not fully controllable and so the task of calibrating a simulator will seldom be possible in cases where $\|GG^+ - \openone\|\not\approx 0$ irrespective of the method used to control it.

It is difficult to say in general when the conditions
underlying~\eq{efficiency} will be met, as it is always possible
for experiments to be chosen that provide virtually no information about the
system.  For example, the observable could be chosen to commute with the
dynamics so that no information can be learned from the measurement statistics.  Great experimental care must be taken in order to ensure that
such pathological cases do not emerge~\cite{wiebe_quantum_2014-1}. These
pathological experiments can be avoided for Ising models with exponential decaying
interactions and we expect exponential decay of $\|\mathcal{E}_{\max}\|$ to be
common for a wide range of models that also include noise and non--commuting terms based on previous
studies~\cite{granade_robust_2012,wiebe_hamiltonian_2014,wiebe_quantum_2014-1}.

The bootstrapped simulator also need not have as many controls as the simulator that is used to certify it.  This does not necessarily mean that the controls in the bootstrapped device are less rich than that of the trusted simulator.  If we assume, for example, that a general Ising simulator is used to bootstrap an Ising simulator with only nearest neighbor couplings (and universal single qubit control) then more general couplings terms can be simulated using two body interactions.  For example, next--nearest neighbor interactions can be simulated using nearest neighbor couplings and single qubit control via:
\begin{align}
&e^{-2iZ\otimes 1\otimes Z \Delta t^2}\ket{\phi}+O(\Delta t^3)\nn
&=e^{-iZ\otimes X\otimes 1 \Delta t}e^{-i1\otimes Y\otimes Z \Delta t}e^{iZ\otimes X\otimes 1 \Delta t}e^{i1\otimes Y\otimes Z \Delta t}\ket{\phi},
\end{align}
where the middle qubit in $\ket{\phi}$ is set to $\ket{0}$. 
Higher--order and parallel methods also exist for engineering such interactions~\cite{borneman_parallel_2012,childs_product_2013}.

\section{Numerical Results}\label{sec:numerical}

In order to show that the compressed quantum Hamiltonian learning and bootstrapping algorithms are scalable
to large systems, we provide numerical evidence that $50$ qubit Ising Hamiltonians with
exponentially decaying interactions can be learned using an $8$ qubit simulator.  We further observe that only a few kilobits of experimental data
are needed to infer an accurate model and that the observable, $A$, that is used for the inference only needs to be supported on a
small number of qubits.  Finally, we apply the compressed quantum Hamiltonian learning algorithm to use the $8$ qubit
simulator to bootstrap a $50$ qubit quantum simulator from an initially uncalibrated device with crosstalk on the controls.  The
bootstrapping procedure reduces the calibration errors in a $50$ qubit simulator by two orders of magnitude using roughly $750$ kilobits of experimental data.  This
calibrated $50$ qubit simulator could then be used to bootstrap an even larger quantum device.

We perform numerical simulations using the open-source QInfer, SciPy and
\texttt{fht} libraries \cite{granade_qinfer:_2012,jones_scipy:_2001,barbey_github:nbarbey/fht_2010}.
All numerical results simulate
using interactive quantum likelihood evaluation
(\fig{qhl-exp-design}) with compressed simulation on an
eight-qubit register.

\subsection{Compressed Quantum Hamiltonian Learning}

Since quantum devices capable of implementing our bootstrapping protocol are not currently available, we examine systems that can be simulated efficiently using classical computers in order to demonstrate that our algorithm applies to large systems. Thus, we focus on the example of an Ising model on a linear chain of qubits, with exponentially decaying interactions,
\begin{equation}
    \label{eq:ising-model}
    H(\vec{x}) = \sum_{i\ne j} x_{i,j} \sigma_z^{(i)} \sigma_z^{(j)},
\end{equation}
where the parameters $x_{i,j}$ are distributed according to the uniform
distribution $x_{i,j}\sim {\rm unif}(0,1)10^{-2(|i-j|-1)}$.  In all cases, the
observable used is $A=(\ketbra{+}{+})^{\otimes a}$ for $a=\{2,4,6\}$, as this
observable is maximally informative for Ising models. For more general models,
a pseudorandom input state and observable can be used instead
\cite{wiebe_quantum_2014-1}.


\fig{experr} shows that a compressed quantum simulator using only $8$ quantum bits is capable of learning a Hamiltonian model for a system with $50$ qubits.
The errors, as measured by the operator norm of difference between the actual Hamiltonian and the inferred Hamiltonian, are typically on the order of $10^{-2}$ after as few as $300$ experiments per scan where $49$ scans are used in total.  This is especially impressive after noting that this constitutes roughly $750$ kilobits of data and that this error of $10^{-2}$ is spread over $1225$ terms.  The data also shows evidence of exponential decay of the error, which is expected from prior studies~\cite{wiebe_hamiltonian_2014,wiebe_quantum_2014-1}.

An important difference between this result and existing QHL schemes~\cite{wiebe_quantum_2014-1,wiebe_hamiltonian_2014} is that the observable will need to be, in some cases, substantially smaller than the simulator.  Choosing a small observable is potentially problematic because it becomes more likely that an erroneous outcome will be indistinguishable from the initial state.  Also, if $a$ is too small then important long-range couplings can be overlooked because their effect becomes hard to distinguish from local interactions.  We find in~\tab{adata} that the cases where $a=4$ and $a=6$ are virtually indistinguishable whereas the median errors are substantially larger for $a=2$, but not substantially worse than $a=4$ for $200$ experiments/scan.  This provides evidence that small $a$ can suffice for Hamiltonian learning.

\begin{table}
\begin{tabular}{|c|c|c|c|}
\hline 
$a$ &$75^{\rm th}$ percentile & Median error & ${25^{\rm th}}$ percentile\\
\hline
$6$ &$0.0043$& $0.0029$ &$0.0014$\\
$4$ &$0.0029$& $0.0018$ & $0.0014$\\
$2$ &$0.0252$& $0.0234$ & $0.0225$\\
\hline
\end{tabular}
\caption{$|\vec{x}-\vec{x}_{\rm true}|_2$ for QHL using different number of qubits for the observable at $500$ experiments/scan.}\label{tab:adata}
\end{table}

\begin{figure*}
  \centering

  \begin{minipage}{0.45\linewidth}
  \includegraphics[width=0.95\textwidth]{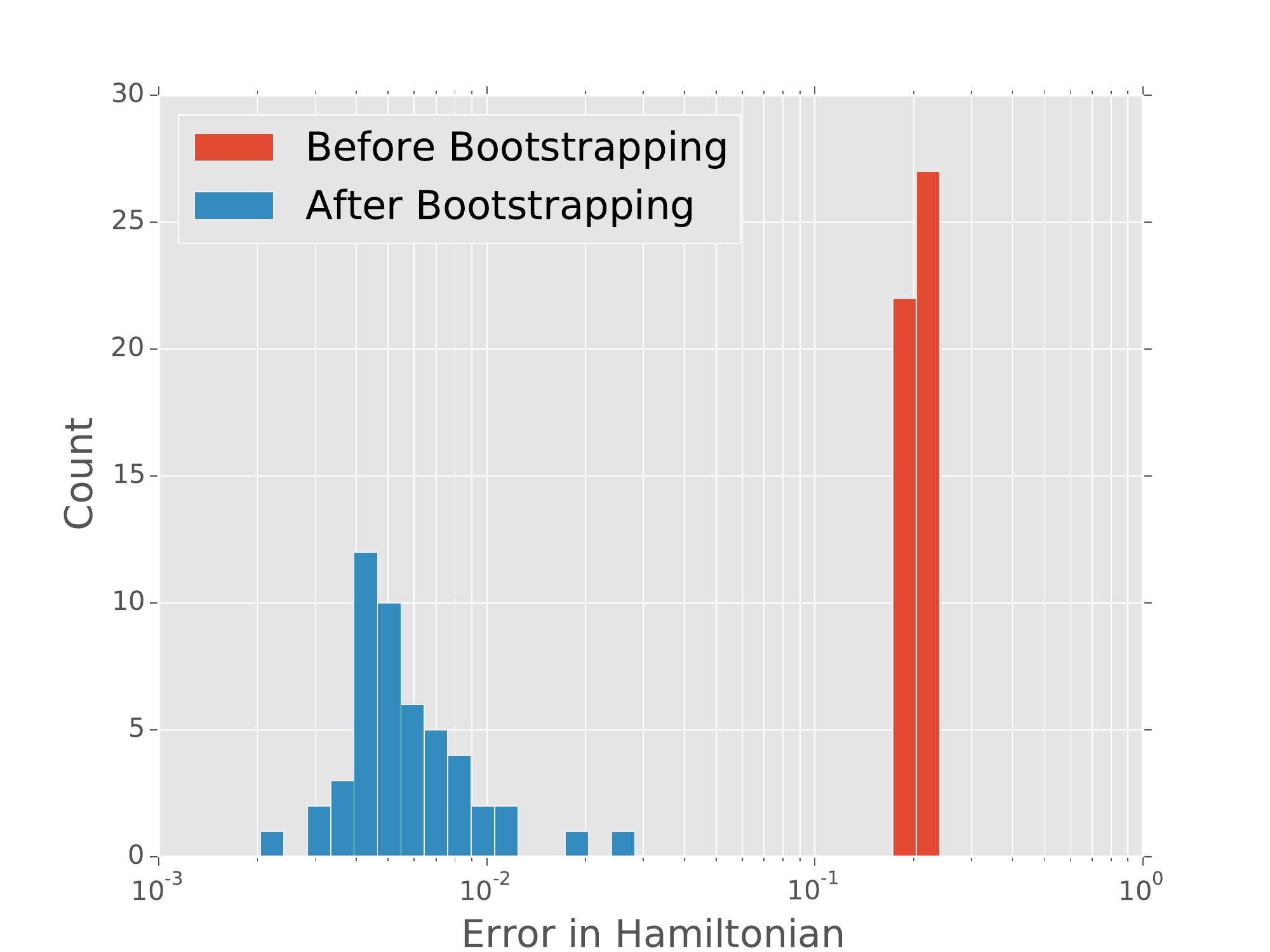}
  \end{minipage}
  \hspace{0.5cm}
  \begin{minipage}{0.45\linewidth}
  \includegraphics[width=0.95\textwidth]{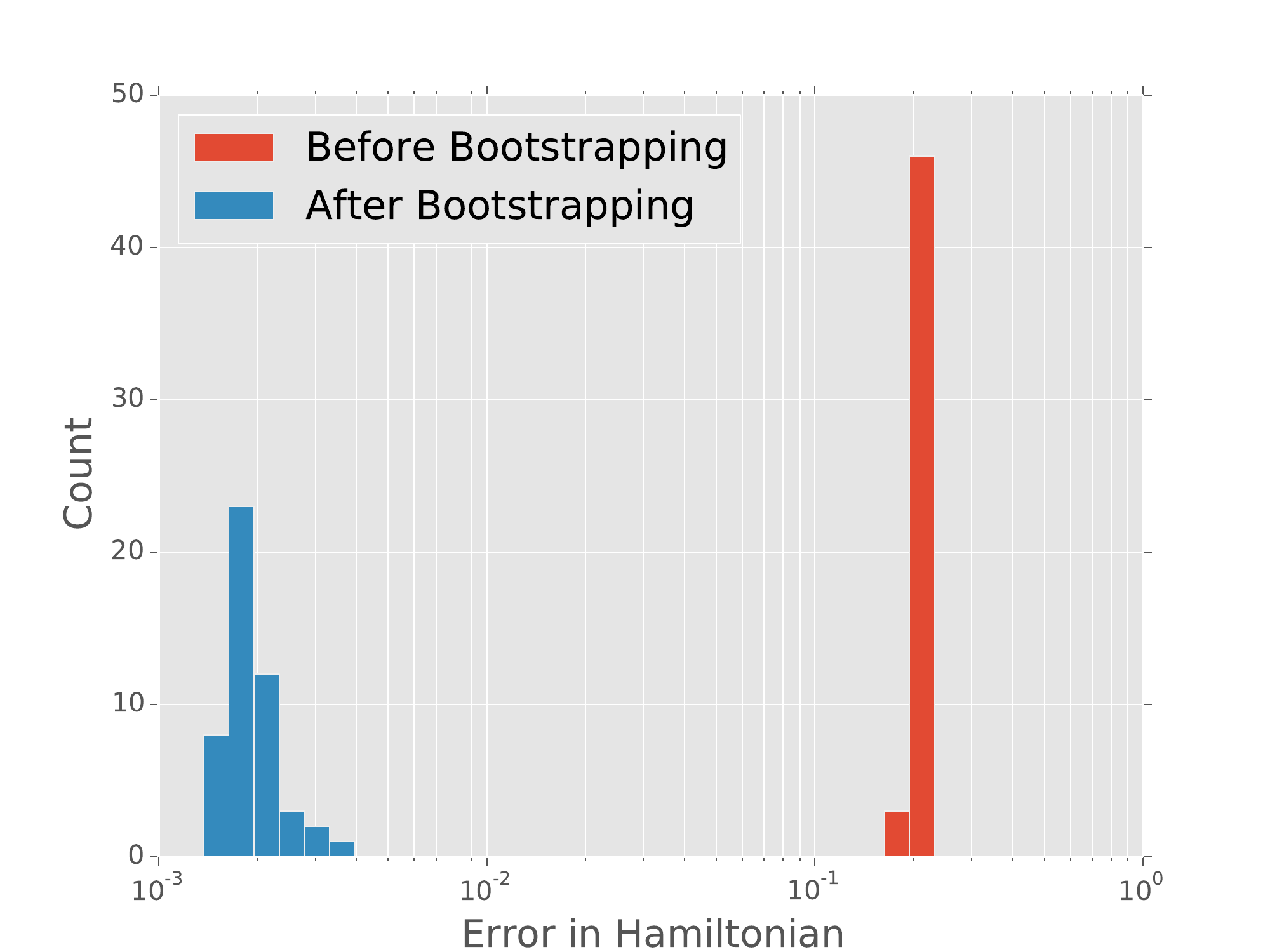}
  \end{minipage}

  \begin{minipage}{0.45\linewidth}
  \includegraphics[width=0.95\textwidth]{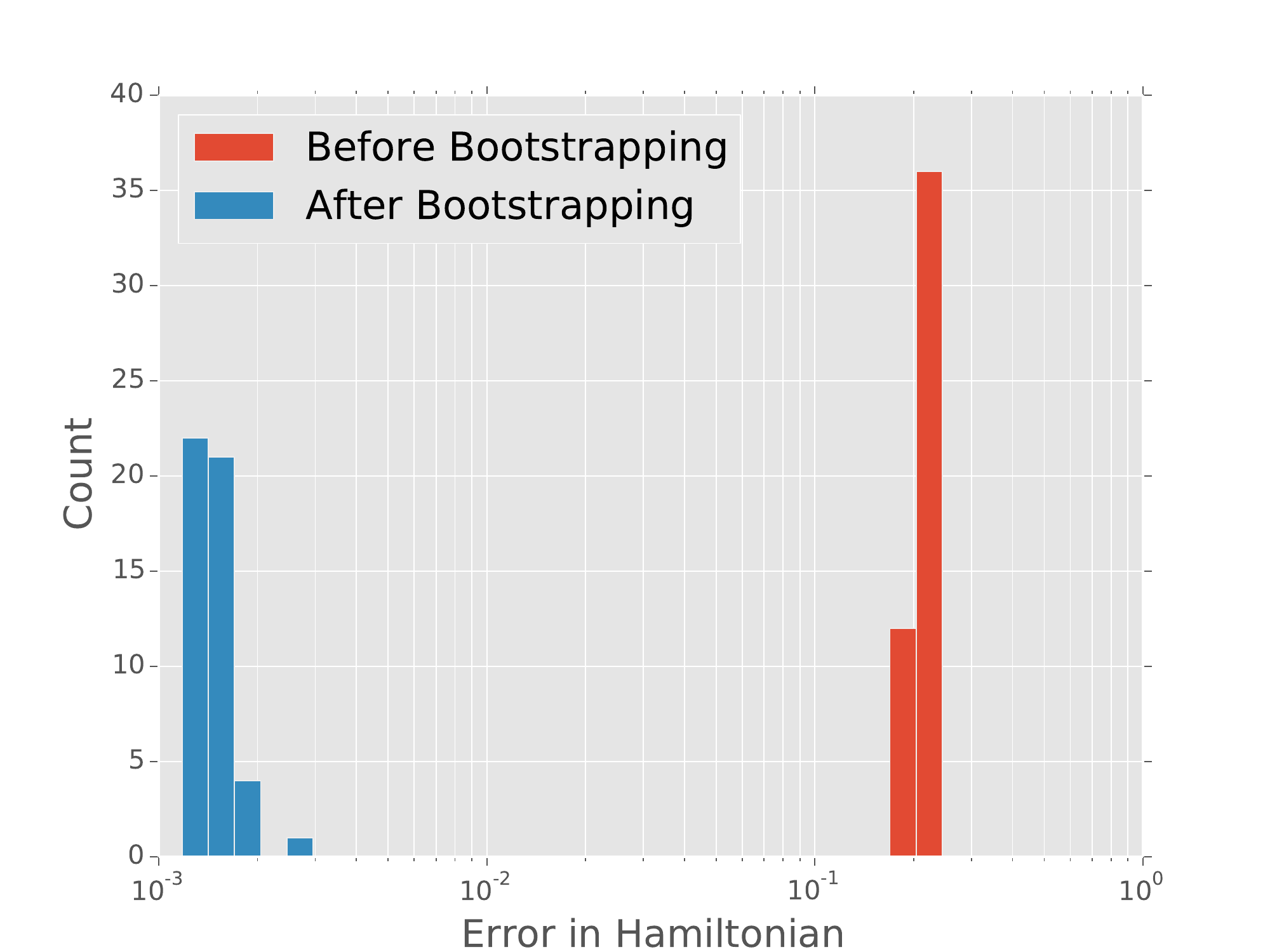}
  \end{minipage}
  \hspace{0.5cm}
  \begin{minipage}{0.45\linewidth}
Errors in inferred control maps\\
  \begin{tabular}{|c|c|c|}
\hline
 Experiments/scan&$\|G - G_{\rm approx}\|_2$&$\|G - G_{\rm prior}\|_2$\\
\hline
$100$&$0.3430$&$2.5625$\\
$200$&$0.0518$&$1.2689$\\
$300$&$0.0323$&$1.2872$\\
\hline
\end{tabular}
\\
~
\\
Errors in couplings before and after bootstrapping\\
  \begin{tabular}{|c|c|c|}
\hline
 Experiments/scan&$\text{Before }$&$\text{After }$\\
\hline
$100$&$0.21\pm 0.01$&$(6.0\pm 4.0)\times 10^{-3}$\\
$200$&$0.21\pm 0.01$&$(2.0\pm 0.4)\times 10^{-3}$\\
$300$&$0.21\pm 0.02$&$(1.5\pm 0.3)\times 10^{-3}$\\
\hline
\end{tabular}
  \end{minipage}
  
  \caption{\label{fig:qhl-hist}
      Distribution of errors for each of the $49$ Hamiltonian terms in the bootstrapped Hamiltonian for a $50$ qubit Ising model using (left)
      $100$ (right) $200$ and (bottom left) $300$ IQLE experiments per scan.
  }
\end{figure*}

\subsection{Quantum Bootstrapping}

 The next set of results show that compressed QHL can be used to bootstrap a quantum simulator for a $50$ qubit $1$D Ising model.  The bootstrapping problem that we consider can be thought of as correcting crosstalk in the large simulator.  This crosstalk manifests itself when the experimentalist attempts to turn on only one of the Ising couplings in the simulator but in fact all $1225$ interactions are actually activated to some extent.    We further assume that the $50$ qubit simulator only has $49$ controls corresponding to each of the nearest neighbor interactions.    This means that a perfect control sequence will generally not exist because $49<1225$.  The control Hamiltonians $[H_1,\ldots,H_M]$ in~\eq{Hvec} conform to~\eq{ising-model} with $x_{i,j}\sim 10\delta_{p,i}\delta_{p,j-1}+{\rm unif}(0,1)10^{-2(|i-j|-1)}$. We also take $H_0=0$.

\fig{qhl-hist} reveals that our bootstrapping procedure reduces control errors by two orders of magnitude in cases where $300$ experiments/scan are used in the QHL step.  Further reductions could be achieved by increasing the number of experiments/scan, but at $300$ scans much of the error arises from $\|GG^+ -\openone\|\ne 0$ so a richer set of controls in the $50$ qubit system would be needed to substantially reduce the residual control errors.  The errors are sufficiently small, however, that it is reasonable that the device could be used as a trusted simulator for nearest--neighbor Ising models. This means that it could be subsequently used to bootstrap another quantum simulator.


\subsection{Scaling with $n$}
All of the examples considered so far examine compressed QHL for $50$ qubits.  Although the fact that the protocol scales successfully up to $50$ qubits already provides strong evidence for its scalability, we provide further evidence here that the errors in compressed QHL do not rapidly vary as a function of the number of qubits in the untrusted system, $n$.  As per the previous numerical examples, we consider a $1$D Ising model with $x_{i,j}\sim {\rm unif}(0,1)10^{-2(|i-j|-1)}$ and use a $4$ qubit observable.  Also $20,000$ particles are used in the SMC approximation to the posterior and we take all data using $200$ experiments per scan.  Roughly $20$ data points per value of $n$ were considered.

\begin{figure}
    \includegraphics[width=0.95\linewidth]{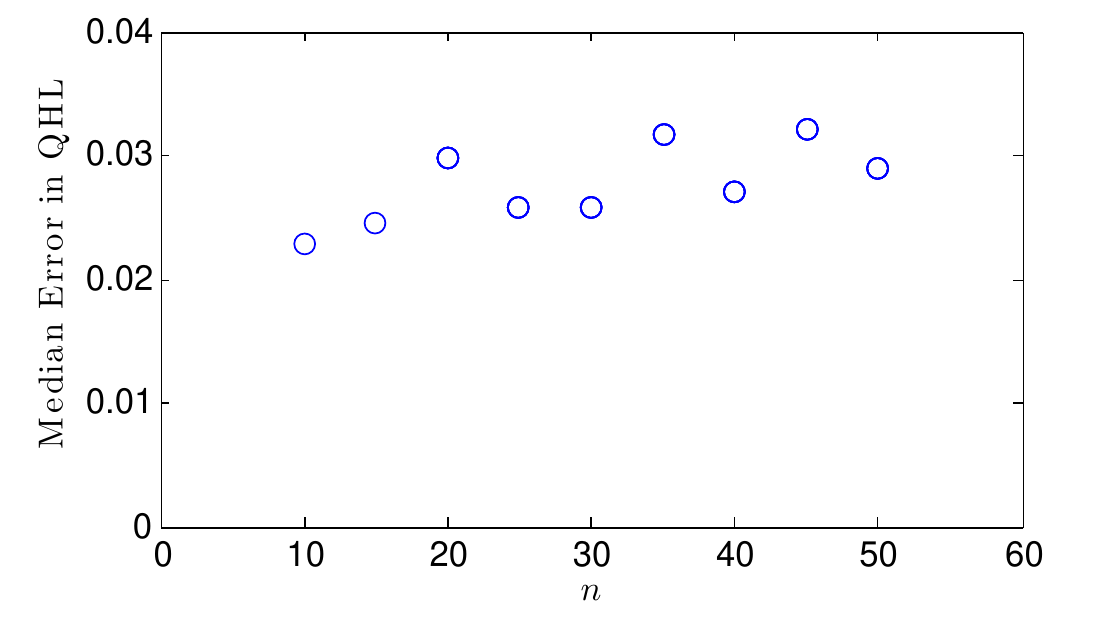}
    \caption{
        Median error in compressed QHL as a function of the number of qubits in the model.
    }
    \label{fig:nscale}
\end{figure}

We see in~\fig{nscale} that the error, as measured by the median ${\rm L}_2$ distance between the inferred model and the true model, is a slowly increasing function of $n$.  The data is consistent with a linear scaling in $n$, although the data does not preclude other scalings.    This suggests that the error in compressed QHL does not rapidly increase for the class of Hamiltonians considered here and provides evidence that examples with far more than $50$ qubits are not outside the realm of possibility for compressed QHL.

\section{Conclusions}

We show that small quantum simulators can be used to characterize
and calibrate larger devices, thus providing a way
to bootstrap to capabilities beyond what can be implemented classically.
In particular, we
provide a compressed quantum Hamiltonian learning algorithm that can infer Hamiltonians for systems with local or rapidly decaying interactions.
The compressed algorithm is feasible because of the fact that local observables remain confined to light cones.  Typically these light cones spread
at a velocity that is dictated by the Hamiltonian. By contrast, in compressed QHL,
light cones spread at a speed that depends on the uncertainty in the Hamiltonian in compressed QHL.
This not only allows more informative experiments to be chosen but also shows that an epistemic speed of light can exist in systems that interact with an intelligent agent.

We then show that this algorithm provides the tools necessary to
bootstrap a quantum system; wherein a small simulator to learn controls that
correct Hamiltonian errors and uncertainties present in a larger quantum device. 
This protocol is useful, for instance, in calibrating control designs to deal with
cross-talk, uncertainties in coupling strengths and other effects that cause
the controls to act differently on the quantum system than the designed
behavior.

Our approaches, being based on quantum Hamiltonian learning, inherit
the same noise and sample error robustness observed in that
algorithm~\cite{wiebe_hamiltonian_2014,wiebe_quantum_2014-1}.
We have provided numerical evidence that our techniques apply to systems with as many
as $50$ qubits, can further tolerate low precision observables, and are
surprisingly efficient. Thus, quantum bootstrapping
provides a potentially scalable technique for application in even
large quantum devices, and in experimentally-reasonable contexts.
Our work therefore provides a critical resource for building practical
quantum information processing devices and computationally useful
quantum simulators.

There are several natural extensions to our work.
While we have focused on the case of time-independent quantum controls and Hamiltonians, our
approaches can be generalized to the time dependent case using more general Lieb--Robinson bounds~\cite{kliesch_lieb-robinson_2014}.
This is significant because techniques such as the method of Da Silva et al~\cite{da_silva_practical_2011} do not apply for $H(t)$.  Additionally,
it would be very interesting to see if quantum simulation can be used to allow local optimization to design even better experiments than those
yielded by the particle guess heuristic.  The introduction of cost effective local optimization strategies may lead to significant advantages
for bootstrapping systems with non--commuting Hamiltonians.


As a final remark, our work provides an important step towards a practical general method for calibrating and controlling large quantum devices,
by utilizing epistemic light cones to compress the simulation, thus enabling
the application of small quantum devices as a resource.
In doing so, our approach also provides a \emph{platform} for building
tractable solutions to more complicated design problems through the application
of quantum simulation algorithms and characterization techniques.



\begin{acknowledgements}
We thank Troy Borneman and Chris Ferrie for suggestions and discussions. CG was supported by funding from Industry Canada, CERC, NSERC, and the Province of Ontario.
\end{acknowledgements}



\ifthenelse{\boolean{IncludeSI}}{

  \onecolumngrid
  \appendix

\label{app:detailed-algorithm}


\section{Fisher Information for Hamiltonian Learning}
\label{app:fisher}

In the main body, we stated that short--time experiments typically do not lead to good estimates of the Hamiltonian parameters.
Here, we justify this claim here by computing the Fisher information, which allows us to
estimate the scaling of the Cram\' er--Rao bound~\cite{van2007parameter}, which lower bounds the
expected variance of any unbiased estimator of the Hamiltonian parameters. 
In particular, the Fisher information matrix can be written for a Hamiltonian
$H = H(\vec{x})$ and measurement in a basis $\{\ket{1}, \dots, \ket{D}\}$ as
\begin{equation}
    \label{eq:fi-defn}
    \matr{I}(H) \defeq \expect_{d} [\vec{\nabla} \ln \Pr(d | H) \vec{\nabla}^\T \ln \Pr(d | H)],
\end{equation}
where $d$ is a random variable representing the outcome of the measurement.

Applying the chain rule and writing out the expectation value gives that
\begin{equation}
    \label{eq:fi2}
    I_{i,j}(H) = \sum_{d\in \{1, \dots, D\}}\frac{\partial_{x_i}{\Pr(d | H)}\partial_{x_j}{\Pr(d | H)}}{\Pr(d | H)}.
\end{equation}
By Born's rule, if we let $U$ be the time--evolution operator for the experiment and define the initial state to be $\ket{0}$ that
\begin{equation}
  \Pr(d | H) = |\braket{d | U | 0}|^2 = \bra{0}U^\dagger \ketbra{d}{d}U\ket{0}.
\end{equation}
We then have that
\begin{equation}
\partial_{x_i} \Pr(d | H)=\bra{0}\partial_{x_i}U^\dagger \ketbra{d}{d}U\ket{0}+\bra{0}U^\dagger \ketbra{d}{d}\partial_{x_i}U\ket{0}.
\end{equation}
Upon substituting back into \eq{fi2}, this yields
\begin{align}
I_{i,j}(H) &= \sum_d \frac{\bra{0}\partial_{x_i}U^\dagger \ketbra{d}{d}U\ket{0}\bra{0}\partial_{x_j}U^\dagger \ketbra{d}{d}U\ket{0}+\bra{0}\partial_{x_i}U^\dagger \ketbra{d}{d}U\ket{0}\bra{0}U^\dagger \ketbra{d}{d}U\ket{0}}{\bra{0}U^\dagger \ketbra{d}{d}\partial_{x_j}U\ket{0}}\nonumber\\
&\qquad+\sum_d \frac{\bra{0}U^\dagger \ketbra{d}{d}\partial_{x_i}U\ket{0}\bra{0}\partial_{x_j}U^\dagger \ketbra{d}{d}U\ket{0}+\bra{0}U^\dagger \ketbra{d}{d}\partial_{x_i}U\ket{0}\bra{0}U^\dagger \ketbra{d}{d}\partial_{x_j}U\ket{0}}{\bra{0}U^\dagger \ketbra{d}{d}U\ket{0}}\nonumber\\
&=\sum_d \bra{0}\partial_{x_i}U^\dagger \ket{d}\bra{0}\partial_{x_j}U^\dagger \ket{d}\frac{\bra{d}U^\dagger \ket{0}}{\bra{0}U \ket{d}}+\bra{0}\partial_{x_i}U \ket{d}\bra{0}\partial_{x_j}U \ket{d}\frac{\bra{0}U \ket{d}}{\bra{d}U^\dagger \ket{0}}\nonumber\\
&\qquad+\sum_d \bra{0}\partial_{x_i}U^\dagger \ket{d}\bra{0}\partial_{x_j}U \ket{d}+ \bra{0}\partial_{x_i}U \ket{d}\bra{0}\partial_{x_j}U^\dagger \ket{d}.\label{eq:Fisher}
\end{align}
It is then straight forward to see that there exists $\phi_d$ such that for every $d$
\begin{equation}
{\bra{0}U \ket{d}}=e^{i\phi_d}{\bra{d}U^\dagger \ket{0}}.
\end{equation}
Furthermore from differentiating $UU^\dagger = \id $ and using the fact that $U$ is unitary, it is clear that for $\|\cdot \|$ the induced $2$--norm, 
\begin{equation}
\|\partial_{x_i} U^\dagger \|=\|\partial_{x_i} U \|.\label{eq:normeq}
\end{equation}
Seeking an upper bound on the Fisher information, we use the Cauchy--Schwarz inequality to show that
\begin{equation}
\sum_d \bra{0}\partial_{x_i}U^\dagger \ket{d}\bra{0}\partial_{x_j}U^\dagger e^{-i\phi_d}\ket{d} \le \sqrt{\sum_d \bra{0}\partial_{x_i} U^\dagger \ketbra{d}{d}\left(\partial_{x_i} U^{\dagger}\right)^\dagger \ket{0}\sum_d \bra{0}\partial_{x_j} U^\dagger \ketbra{d}{d}\left(\partial_{x_j} U^{\dagger}\right)^\dagger \ket{0}}.\label{eq:cauchy}
\end{equation}
Using the resolution of unity and the fact that $\|A^\dagger\|=\|A\|$ for the $2$--norm, we find from~\eqref{eq:cauchy} and~\eqref{eq:normeq} that
\begin{equation}
\sum_d \bra{0}\partial_{x_i}U^\dagger \ket{d}\bra{0}\partial_{x_j}U^\dagger e^{-i\phi_d}\ket{d} \le \|\partial_{x_i} U\|\|\partial_{x_j} U\|.\label{eq:cauchy2}
\end{equation}
The triangle inequality and equations~\eqref{eq:cauchy2} and~\eqref{eq:Fisher} then imply that
\begin{equation}
I_{i,j}(H)\le 4\|\partial_{x_i} U\|\|\partial_{x_j} U\|.
\end{equation}

An experiment for either the case where an inversion step is employed or the case where only forward
evolution is used can be written using the unitary $U=e^{i\Hinv t}e^{-iHt}$, where $\Hinv=0$ for
the inversion--free case.
Regardless, $\Hinv$ is explicitly independent of the parameters $\{x_p\}$ of $H$;  therefore since $U$ is unitary,
\begin{equation}
\|\partial_{x_p} U\|=\|e^{i\Hinv t} \partial_{x_p}e^{-iHt}\| = \|\partial_{x_p}e^{-iHt}\|.
\end{equation}
Using the definition of the parametric derivative of an operator exponential, we find using the triangle inequality that
\begin{equation}
\|\partial_{x_p}e^{-iHt}\|= \left\|\int_0^1 e^{(1-\tau)(-iHt)}(-it\partial_{x_p} H)e^{\tau(-iHt)}\mathrm{d}\tau \right\|\le \|\partial_{x_p} H\|t.
\end{equation}
This leads us to the conclusion that
\begin{equation}
I_{i,j}(H)\le 4\|\partial_{x_i} H\|\|\partial_{x_j} H\| t^2.\label{eq:Fisherbd}
\end{equation}
The Cram\' er--Rao bound then states that, for any unbiased estimator $\hat{\vec{x}}$
of the Hamiltonian parameters \cite{cover_elements_2006},
\begin{equation}
    \expect_{\text{data}} [\Cov(\hat{\vec{x}})] - \matr{I}^{-1}(\vec{x}) \ge 0,
\end{equation}
where the expectation value is taken over all data records, here taken to
be measurements of $d\in\{1, \dots, D\}$~\cite{van2007parameter}. Tracing both sides of the inequality immediately
implies that the variance of any unbiased estimator of the
Hamiltonian parameters scales with $\Tr[\matr{I}(H)^{-1}] \in \Omega(t^{-2})$.

We also consider the Bayesian Cram\'er-Rao bound \cite{van_trees_detection_1968,gill_applications_1995,dauwels_computing_2005},
which bounds the performance of biased estimators by taking the expectation
of the Cram\'er-Rao bound over a prior $\pi$,
\begin{equation}
    \expect_{\vec{x} \sim \pi} [\expect_{\text{data}} [\Cov(\hat{\vec{x}})]] - \expect_{\vec{x} \sim \pi} [\matr{I}^{-1}(H(\vec{x}))] \ge 0.
\end{equation}
Here, we note that the $t^2$ scaling obtained in \eq{Fisherbd} is
independent of $\vec{x}$, it factors out of the expectation over
Hamiltonian parameters, such that the Bayesian Cram\'er-Rao bound
is also $\Omega(t^{-2})$ by the same argument, such that even biased
estimators require evolution time that is the reciprocal of the desired standard
deviation.

This implies that as $t\rightarrow 0$ the lower bound on the variance of the
optimal estimator for $\vec{x}$ diverges, implying that the experiments become
uninformative for the small values of time required for existing Hamiltonian
identification methods to succeed.  Also, since the cost of performing an
experiment becomes dominated by the time required to prepare the initial state
for small $t$, it is clear that the reduced cost of short--time experiments will
not compensate for the exponentially diverging CRB and BCRB in such cases.

\section{Lieb--Robinson Bounds}
\label{app:lrb}
We provide rigorous estimates for the truncation error
in cases where the Hamiltonian is non--commuting in this section.
The proof of the error bounds is elementary, with the exception that the results depend on the use of Lieb--Robinson bounds.
To begin, let us first define some notation.
Let us assume that $r$ time reversals are used, that the Trotter formula (rather than higher order variants) is used, and then let us define
the observable after $n$ evolutions/inversions to be
\begin{equation}
A^{(n)} := e^{iHt/r}e^{-iH_-t/r}A^{(n-1)}e^{iH_-t/r}e^{-iHt/r},
\end{equation}
with $A^{(0)}=A$.
Now, let $H_- = H_{\rm in} - \Lambda$, were $\Lambda$ is the discrepancy between the inversion Hamiltonian and the true
Hamiltonian, supported on the region that can be simulated by the trusted device.  Let us also define
\begin{equation}
\tA{n} := e^{i\Lambda t/r} \tA{n-1} e^{-i\Lambda t/r},
\end{equation}
where $\tA{0} =A$; since we have assumed a Trotter formula, $[H_-, H_{\mathrm{int}}] t / r \approx 0$,
such that $\tA{n}$ represents the observable as simulated by the trusted device alone.
Thus, define the error operator $\delta^{(n)}$ such that
\begin{equation}
 \delta^{(n)}:=A^{(n)}- \tA{n}.
\end{equation}
The goal of this section will then be to provide upper bounds on $\|\delta^{(n)}\|$ which represents the error incurred from truncating the 
trusted simulator.

First, note that $\|\delta^{(0)}\|=0$.  This will serve as the base case in our inductive argument about the norm of $\delta^{(n)}$.  The triangle inequality, together with the unitary invariance of $\|\cdot\|$, implies
\begin{align}
\|\delta^{(n+1)}\| &= \| e^{iHt/r}e^{-iH_-t/r}A^{(n)}e^{iH_-t/r}e^{-iHt/r} - e^{i\Lambda t/r} \tA{n} e^{-i\Lambda t/r}\|\nonumber\\
& \le \|\delta^{(n)}\|+\| e^{iHt/r}e^{-iH_-t/r}\tA{n}e^{iH_-t/r}e^{-iHt/r} - e^{i\Lambda t/r} \tA{n} e^{-i\Lambda t/r}\|.\label{eq:errrecurse}
\end{align}
Eq.~\eq{errrecurse} provides a recursive expression for the error after $n+1$ steps in terms of the error after $n$ steps.  Our bounds for the error in~\eq{lrberror} follow from unfolding this recurrence relation after   applications of the triangle inequality.  The main challenge is that $e^{iHt/r}e^{-iH_-t/r}\tA{n}e^{iH_-t/r}e^{-iHt/r} - e^{i\Lambda t/r} \tA{n} e^{-i\Lambda t/r}$ is difficult to bound directly.  So instead we introduce a telescoping series of terms such that the difference between any two consecutive terms in the series can be estimated.  We then arrive at~\eq{lrberror} by using the triangle inequality.

Our first such step considers the Trotter error involved in using the approximation $e^{-iH_- t/r}\approx e^{-iH_{\rm in} t/r}e^{-i\Lambda t/r}$ in~\eq{errrecurse}.  
First, note that
\begin{align}
&\|e^{iHt/r}e^{-iH_-t/r}\tA{n}e^{iH_-t/r}e^{-iHt/r} - e^{iHt/r}e^{-iH_{\rm in}t/r}\tA{n+1}e^{iH_{\rm in}t/r}e^{-iHt/r}\|\nonumber\\ &\qquad=\|e^{-iH_-t/r}\tA{n}e^{iH_-t/r} - e^{-iH_{\rm in}t/r}e^{i\Lambda t/r}\tA{n}e^{-i\Lambda t/r}e^{iH_{\rm in}t/r}\|=\|[e^{-i\Lambda t/r}e^{iH_{\rm in}t/r}e^{-iH_-t/r},\tA{n}]\|.\label{eq:commrel1}
\end{align}
Using the result of Huyghebaert and De Raedt \cite{huyghebaert_product_1990}, we have that for any two operators $A,B$ that have commutators of bounded norm,
\begin{equation}
\|e^{A}e^{B} -e^{A+B}\|\le \frac{1}{2}\|[A,B]\|.\label{eq:hdr}
\end{equation}
Then, using~\eq{hdr} we see that there exists an operator $C$ such that $\|C\|\le 1$ and $e^{A}e^Be^{-(A+B)}=1 +  \frac{C}{2}\|[A,B]\|$.  

Now noting that $H_-=H_{\rm in}-\Lambda$, we see that
\begin{align}
\|[e^{-i\Lambda t/r}e^{iH_{\rm in}t/r}e^{-iH_-t/r},\tA{n}]\|& = \left\|\left[e^{-i\Lambda t/r}e^{iH_{\rm in}t/r}e^{-i(H_{\rm in}- \Lambda)t/r},\tA{n}\right]\right\|.\label{eq:commrel2}
\end{align}
Therefore there exists an operator $C$ with norm at most one such that
\begin{align}
\left\|\left[e^{-i\Lambda t/r}e^{iH_{\rm in}t/r}e^{-i(H_{\rm in}- \Lambda)t/r},\tA{n}\right]\right\| &=\left\|\left[1 +\frac{C \left\|\left[H_{\rm in},\Lambda\right]\right\|{t^2}}{2{r^2}},\tA{n}\right]\right\| .\nn
&\le \|[H_{\rm in},\Lambda]\|\|\tA{n}\|{t^2}/{r^2}\nn
&=\|[H_{\rm in},\Lambda]\|\|A\|{t^2}/{r^2}.\label{eq:commsol1}
\end{align}
Eq.~\eq{commsol1} provides an upper bound for the error incurred by treating the evolution on the trusted simulator as if it were evolving separately under $H_{\rm in}$ and $\Lambda$ during the inversion phase, rather than evolving under $H_-= \H_{\rm in} - \Lambda$.

Second, we have from similar reasoning and the facts that (a) $H=H_{\rm out} + H_{\rm int} + H_{\rm in}$ and (b) $H_{\mathrm{in}}$ and $H_{\textrm{out}}$ are disjoint in support and hence $[H_{\rm in}, H_{\rm out}]=0$ that
\begin{equation}
\|e^{iHt/r}e^{-iH_{\rm in}t/r}\tA{n+1}e^{iH_{\rm in}t/r}e^{-iHt/r} - e^{i(H_{\rm out} +H_{\rm int})t/r}\tA{n+1}e^{-i(H_{\rm out} +H_{\rm int})t/r}\|\le \|[H_{\rm int},H_{\rm in}]\|\|A\|t^2/r^2.\label{eq:commsol2}
\end{equation}
It is then straightforward to see from adding and subtracting appropriate terms and then applying the triangle inequality that
\begin{equation}
\|\delta^{(n+1)}\| \le \|\delta^{(n)}\| +(\|[H_{\rm in},\Lambda]\| + \|[H_{\rm int},H_{\rm in}]\|)\|A\|\frac{t^2}{r^2}+\|e^{i (H_{\rm out}+H_{\rm int})t/r}\tA{n+1}e^{-i(H_{\rm out} +H_{\rm int})t/r}-\tA{n+1}\|.\label{eq:delta1}
\end{equation}

Third, as illustrated in \fig{bootstrapping-partition}, there are two types of interaction terms: interactions between the neglected particles and those in the support of $A$ and interactions between neglected qubits and those not in the support of $A$.  The Hamiltonians composed of only these interactions are denoted $H_{{\rm int}\!\bigcap\!A}$ and $H_{{\rm int}\setminus A}$, such that
\begin{align}
&\|e^{i(H_{\rm out} + H_{\rm int})t/r}\tA{n+1}e^{-i(H_{\rm out} + H_{\rm int})t/r} - e^{i(H_{\rm out} +H_{{\rm int}\setminus A})t/r}\tA{n+1}e^{-i(H_{\rm out} +H_{{\rm int}\setminus A})t/r}\|\nonumber\\
&\qquad = \|[e^{-i(H_{\rm out} +H_{{\rm int}\setminus A})t/r}e^{i(H_{\rm out} + H_{\rm int})t/r},\tA{n+1}]\|.\label{eq:A9}
\end{align}
Using the fact that $\|e^{-i(H_{\rm out} +H_{{\rm int}\setminus A})t/r}e^{i(H_{\rm out} + H_{\rm int})t/r}-1\|\le \|H_{{\rm int}\bigcap A}\| t/r$,
the triangle inequality yields
\begin{equation}
\|[e^{-i(H_{\rm out} +H_{{\rm int}\setminus A})t/r}e^{i(H_{\rm out} + H_{\rm int})t/r},\tA{n+1}]\|\le 2\|H_{{\rm int}\bigcap A}\|\|A\| t/r.
\end{equation}
This bound estimates the error incurred by neglecting direct interactions between the observable and the particles omitted from the trusted simulator.
Thus~\eq{delta1} can be simplified to
\begin{align}
\|\delta^{(n+1)}\| &\le \|\delta^{(n)}\| +(\|[H_{\rm in},\Lambda]\| + \|[H_{\rm int},H_{\rm in}]\|)\|A\|\frac{t^2}{r^2}+2\|H_{{\rm int}\bigcap A}\|\|A\| t/r\nonumber\\
&\qquad+\|e^{i (H_{\rm out}+H_{{\rm int}\setminus A})t/r}\tA{n+1}e^{-i(H_{\rm out} +H_{{\rm int}\setminus A})t/r}-\tA{n+1}\|.\label{eq:delta2} 
\end{align}
Using Hadamard's lemma, this can be written as
\begin{align}
\|\delta^{(n+1)}\| &\le \|\delta^{(n)}\| +(\|[H_{\rm in},\Lambda]\| + \|[H_{\rm int},H_{\rm in}]\|)\|A\|\frac{t^2}{r^2}+2\|H_{{\rm int}\bigcap A}\|\|A\| t/r\nonumber\\
&\qquad+\|i[H_{\rm out}+H_{{\rm int}\setminus A},\tA{n+1}] t/r -\frac{1}{2!}[H_{\rm out}+H_{{\rm int}\setminus A},[H_{\rm out}+H_{{\rm int}\setminus A},\tA{n+1}]]t^2/r^2+\cdots\|.\label{eq:delta3}
\end{align}
Applying the triangle inequality, factoring and recombining terms as an exponential yields
\begin{align}
\|\delta^{(n+1)}\| &\le \|\delta^{(n)}\| +(\|[H_{\rm in},\Lambda]\| + \|[H_{\rm int},H_{\rm in}]\|)\|A\|\frac{t^2}{r^2}+2\|H_{{\rm int}\bigcap A}\|\|A\| t/r\nonumber\\
&\qquad+\|[H_{\rm out}+H_{{\rm int}\setminus A},\tA{n+1}]\|e^{2\|H_{\rm out}+H_{{\rm int}\setminus A}\|t/r} t/r\label{eq:delta4}.
\end{align}

Fourth, and finally, we apply the Lieb--Robinson bound to upper bound $\|[H_{\rm out}+H_{{\rm int}\setminus A},\tA{n+1}]\|$.  Assuming that the interactions that comprise $H_{\rm int}$ are nearest--neighbor or exponentially decay with the graph distance between the qubits in question, the Lieb--Robinson bound states that there exist constants $s$ and $\mu$ that are only dependent on the properties of $\Lambda$~\cite{hastings_spectral_2006} such that
\begin{equation}
\|[H_{\rm out}+H_{{\rm int}\setminus A},\tA{n+1}]\|\le 2\|H_{\rm out} + H_{{\rm int}\setminus A}\| \|A\| |\{A\}|e^{-\mu {\rm dist}(A,H_{\rm out})}\left[e^{2s|t|(n+1)/r} -1 \right].\label{eq:lrbeqn}
\end{equation}
Substituting~\eq{lrbeqn} into~\eq{delta4} and noting that $[H_{\rm out}, \tA{n+1}]=0$ (because $\Lambda$ and $H_{\rm out}$ have disjoint support) yields
\begin{align}
\|\delta^{(n+1)}\| &\le \|\delta^{(n)}\| +(\|[H_{\rm in},\Lambda]\| + \|[H_{\rm int},H_{\rm in}]\|)\|A\|\frac{t^2}{r^2}+2\|H_{{\rm int}\bigcap A}\|\|A\| t/r\nonumber\\
&\qquad+2\| H_{{\rm int}\setminus A}\| \|A\| |\{A\}|e^{-\mu {\rm dist}(A,H_{\rm out})}\left[e^{2s|t|(n+1)/r} -1 \right]e^{2\|H_{\rm out}+H_{{\rm int}\setminus A}\|t/r} t/r\label{eq:delta5}.
\end{align}

Applying~\eq{delta5} recursively, it is then clear that
\begin{align}
\|A(t) - \tilde{A}(t)\|=\|\delta^{(r)}\| &\le (\|[H_{\rm in},\Lambda]\| + \|[H_{\rm int},H_{\rm in}]\|)\|A\|\frac{t^2}{r}+2\|H_{{\rm int}\bigcap A}\|\|A\| t\nonumber\\
&\qquad+2\| H_{{\rm int}\setminus A}\| \|A\| |\{A\}|te^{-\mu {\rm dist}(A,H_{\rm out})}\left[e^{2s|t|} -1 \right]e^{2\|H_{\rm out}+H_{{\rm int}\setminus A}\|t/r} \label{eq:delta6}.
\end{align}

There are a few interesting points to note about~\eq{delta6}.  Firstly, the Lieb--Robinson velocity that appears in the equation is that of $\Lambda$ not $H$. If the particle guess heuristic is used to select experiments, then the speed at which the commutator depends on the \emph{uncertainty} in the Hamiltonian rather than the actual Hamiltonian.  A consequence of this is that the Lieb--Robinson velocity here is an epistemic, rather than a physical, property of the system.  This means that, even though the experimental times increase under the particle guess heuristic as more information is learned about $H$, the Lieb--Robinson velocity relevant to this problem will shrink.  
Very long experiments can therefore be used without requiring that the distance between $A$ and the neglected qubits (i.e. the volume of the trusted simulator) grows linearly with the evolution time.  In particular, ${\rm dist}(A,H_{\rm out})$ must grow at most logarithmically with the evolution time rather than linearly.

It should also be noted that in cases where nearest--neighbor couplings are present, rather than exponentially decaying couplings, that these bounds are known to be loose.  More sophisticated treatments of the Lieb--Robinson bounds show that the error shrinks as $e^{-{\rm const}\times  {\rm dist}(A,H_{\rm out})^2}$ for such systems~\cite{hastings2010locality}.  Taken together
with the observation that $H_{{\rm int}\bigcap A}=0$ for nearest neighbor couplings,
since $A$ is not supported on the boundary of the trusted simulator,
this tighter scaling implies that the volume of the trusted simulator can be quadratically smaller in such cases.

\section{Bounds for $1$-D Ising Models}\label{app:scale}
In the main body, we showed that for commuting Hamiltonians (such as the Ising model) that the maximum evolution time allowed by compressed simulation is dictated by the strength of the  interactions between the observable and the neglected subsystems.  Here we provide bounds that are useful for estimating the maximum value of this interaction strength, which is useful for knowing when the simulations fail.  We used these bounds in our numerical simulations to set limits on the allowable evolution time, thereby allowing us to simulate dynamics on the full $50$ qubit system on a desktop.

As particular examples, if we assume that the Hamiltonian is an Ising model on a line of length $\ell$ with non--nearest neighbor couplings between sites $i$ and $j$ that scale at most as $be^{-\alpha|i-j|}$, $A$ is supported on $a$ sites and the trusted simulator can simulate $w$ sites then 
\begin{equation}
\|H_{{\rm int}\bigcap A}\|^{-1}\ge (1-e^{-\alpha})e^{\lfloor\frac{w-a}{2} \rfloor}/ab.
\end{equation}
It therefore suffices to take $w-a$ logarithmic in $t$ to guarantee error of $\delta$ for any fixed $t$.  Similarly, if we assume the interaction strength between sites $i$ and $j$ is at most $b/|i-j|^\alpha$ for $\alpha>1$ then
\begin{equation}
\|H_{{\rm int}\bigcap A}\|^{-1}\ge \frac{(\left \lfloor\frac{w-a}{2} \right \rfloor +1)^\alpha(\alpha-1)}{ab\alpha}.
\end{equation}
Picking $w-a\in O(t^{1/\alpha})$ guarantees fixed error $\delta$ for experimental time $t$.  These scalings are justified below.

Assume that the Hamiltonian is an Ising model on a line of length $\ell$ with a trusted simulator that can simulate at most $w$ sites and an observable that has support on $a$ sites.  We then can write the norm of the Hamiltonian terms that are neglected by the trusted simulator as $\|H_{{\rm int}\bigcap A}\| \le a\sum_{j=\lfloor (w-a)/2\rfloor +1}^{\ell -a} f(j)$ for some function $f(j)$ that describes how quickly the interactions decay with distance from the observable.  Here we take the lower limit of the sum to be $\lfloor (w-a)/2 \rfloor+1$ because this is the closest possible site within the support of the observable $A$ to the un-modeled portion of the spin chain.  Note that in cases of non-periodic boundary conditions this minimum distance may be farther for simulations that occur near the end of the chain.  Similarly, the furthest any site can be in the chain from $A$ is $\ell -a$ which justifies the upper bound for the sum.  Again this upper limit may not be tight for periodic boundary conditions.

The two most interesting cases, experimentally, are cases with exponential decay and polynomial decay with $j$.  If we assume that $f(j)\le be^{-(j-1)\alpha}$ then
\begin{equation}
\sum_{j=\lfloor (w-a)/2\rfloor+1}^{\ell -a}a f(j) \le \sum_{j=\lfloor (w-a)/2\rfloor+1}^\infty abe^{-(j-1)\alpha}= \frac{ab e^{-\lfloor \frac{w-a}{2}\rfloor \alpha}}{1-e^{-\alpha}}.
\end{equation}
This justifies the claim made in the main body.

Polynomial decay is similar.  Assume $f(j)\le b/j^\alpha$ then
\begin{align}
\sum_{j=\lfloor\frac{w-a}{2}\rfloor+1}^{\ell-a} \frac{ab}{j^{\alpha}} &= \frac{ab}{(\lfloor \frac{w-a}{2}\rfloor +1)^\alpha}\sum_{k=0}^{\ell-a-\lfloor\frac{w-a}{2} \rfloor-1}\frac{1}{(1+\frac{k}{(w-j)/2})^\alpha}\nonumber.\\
&\le\frac{ab}{(\lfloor \frac{w-a}{2}\rfloor +1)^\alpha}\left(1+\int_{0}^{\ell-a-\lfloor\frac{w-a}{2} \rfloor-1}\frac{1}{(1+\frac{k}{(w-j)/2})^\alpha}\mathrm{d}k\right).
\end{align}
This bound can be evaluated for cases where $\alpha=1$, and logarithmic divergence with $\ell$ will be observed in those cases.  Given the assumption that $\ell>1$, the integral is convergent so for simplicity we can take the limit of this equation as $\ell\rightarrow \infty$.  Evaluating the integral and some elementary simplifications leads t
\begin{equation}
\sum_{j=\lfloor\frac{w-a}{2}\rfloor+1}^{\ell-a} \frac{ab}{j^{\alpha}} \le \frac{ab\alpha}{(\lfloor \frac{w-a}{2}\rfloor +1)^\alpha(\alpha-1)}.
\end{equation}
This justifies the claim in the main body about polynomial scaling and shows that increasing $w$ to increases the maximum value of $t$ allowable in the experiment design step.

\section{Bounds for Errors in Bootstrapping}\label{app:booterror}
To begin let us consider the error incurred by trying to find a control sequence that produces a Hamiltonian $\mathcal{H}_k$ on an initially untrusted quantum device.  If the inferred control map is $G_1$ and the actual control map is $G_1 +\mathcal{E}_1$ then the error in the implemented Hamiltonian, after one bootstrapping step, is
\begin{equation}
\|((G_1 + \mathcal{E}_1)G_1^+ -\openone)\mathcal{H}_k\| \le (\|G_1G_1^+ -\openone\| + \|\mathcal{E}_1\|\|G^+\|)\|\mathcal{H}_k\|.
\end{equation}

Now let us consider the error incurred after bootstrapping $L$ times.  Or in other words, consider the error that arises from using a trusted simulator that was calibrated via $L-1$ steps of bootstrapping.  If we define $G_j$ and $\mathcal{E}_j$ to be the control maps and error operators that arise after $j$ steps (where each $\mathcal{E}_j$ is the error with respect to the ``trusted simulator'' calibrated via $j-1$ bootstrapping steps) then the error is
\begin{equation}
\|(((G_L +\mathcal{E}_L)G^+_L)(G_{L-1}+\mathcal{E}_{L-1})G_{L-1}^+\cdots (G_{1}+\mathcal{E}_{1})G_1^+-\openone) \mathcal{H}_k\|.\label{eq:errgen}
\end{equation}
By adding and subtracting $(G_1+\mathcal{E}_1)G^+_1$, $(G_2+\mathcal{E}_2)G^+_2(G_1+\mathcal{E}_1)G^+_1$ and so forth from~\eq{errgen} we obtain from the triangle inequality that the error is at most
\begin{equation}
\sum_{j=1}^L (\|G_jG_j^+ -\openone\|+\|\mathcal{E}_j\|\|G_j^+\|)\prod_{k=1}^{j-1} \left(\|G_k\|\|G_k^+\|+ \|\mathcal{E}_k\|\|G_k^+\|\right).\label{eq:totalerr}
\end{equation}
Noting that the condition number, $\kappa_k$, for $G_k$ is $\|G_k\|\|G^+_k\|$,~\eq{totalerr} can be upper bounded by the maximum values of of each of the terms involved.  If we specifically define $\Gamma_{\max}$ to be the maximum value of $\|G_jG_j^+ -\openone\|+\|\mathcal{E}_j\|\|G_j^+\|$ and $\kappa_{\max}$ to be the maximum condition number then
\begin{equation}
\sum_{j=1}^L (\|G_jG_j^+ -\openone\|+\|\mathcal{E}_j\|\|G_j^+\|)\prod_{k=1}^{j-1} \left(\|G_k\|\|G_k^+\|+ \|\mathcal{E}_k\|\|G_k^+\|\right)\le L\Gamma_{\max}(1+(\kappa_{\max}-\openone)+\|\mathcal{E}_{\max}\|\|G^+_{\max}\|)^{L-1}. \label{eq:totalerr2}
\end{equation}
The result in~\eq{errscale} then follows from the fact that $(1+x) \le e^x$ for all $x\in\mathbb{R}$.

Note that this bound is expected to be quite pessimistic for bootstrapping in general.  The analysis makes liberal use of the triangle inequality and uses worst case estimates on top of that.  Additionally, the user in the bootstrapping protocol has some knowledge of the error from the fact that $G_jG^+ -\openone$ can be computed for these problems since the matrices are of polynomial size.  We avoid including this knowledge in the argument since the user does not necessarily know what $\mathcal{E}_j$ is and hence it is conceivable in extremely rare cases that the errors from the approximate inversion could counteract the errors in the Hamiltonian inference.  A more specialized argument may be useful for predicting better bounds for the error in specific applications.

Finally, if the swap gates also have miscalibration errors of $\Delta$ then there is a maximum value of $r$ that can be used before the contributions of such errors become dominant.  A simple inductive argument shows that 
\begin{equation}
r\le \frac{1}{2} \left(\frac{\delta}{\Delta} +1 \right),
\end{equation} 
suffices to guarantee that such errors sum to at most $\delta$.  This shows that the protocol is only modestly sensitive to such errors and that if quantum bootstrapping is used to calibrate the swap gates then it is reasonable to expect that $\delta$ can often be made sufficiently small using a logarithmic number of experiments.
}{} 
%

\end{document}